%% file: main.tex
\RequirePackage{fix-cm}
\documentclass[twocolumn]{svjour3}          
\smartqed  
\makeatletter
\def\cl@chapter{\@elt {theorem}}
\makeatother
\usepackage{graphicx}
\usepackage{mathptmx}      
\usepackage[switch,columnwise]{lineno}
\usepackage{pgfplots}
\usepackage{algorithm}
\usepackage{algpseudocode}
\usepackage{pifont}
\usepackage{mwe}
\usepackage{subfig}
\usepackage{caption}
\usepackage[capitalise]{cleveref}
\usepackage{enumitem}   
\usepackage{array}
\usepackage{tikz}
\usetikzlibrary{pgfplots.groupplots}
\pgfplotsset{compat=newest}
\usepackage{array} 
\usepackage{pgfplotstable}
\usepgfplotslibrary{groupplots}
\usepgfplotslibrary{fillbetween}
\usepackage{url}
\usepackage{tabularx}
\usepackage{gensymb}
\newcolumntype{s}{>{\hsize=.5\hsize}X}

\pgfplotsset{compat=1.5}
\definecolor{bblue}{HTML}{0000CD}
\begin{document}
\title{A topology-oblivious routing protocol for NDN-VANETs}  



\author{Eirini Kalogeiton \and
       Thomas Kolonko \and
       Torsten Braun 
}

\institute{Eirini Kalogeiton \at University of Bern, Institute of Computer Science, Bern, Switzerland
              \email{kalogeiton@inf.unibe.ch}         
           \and
          Thomas Kolonko \at University of Bern, Institute of Computer Science, Bern, Switzerland
             \email{thomas.kolonko@students.unibe.ch}
           \and
           Torsten Braun \at University of Bern, Institute of Computer Science, Bern, Switzerland
           \email{braun@inf.unibe.ch}
}

\date{Received: date / Accepted: date}

\maketitle

\begin{abstract}
Vehicular Ad Hoc Networks (VANETs) are characterized by intermittent connectivity, which leads to failures of end to end paths between nodes.
Named Data Networking (NDN) is a network paradigm that deals with such problems, since information is forwarded based on content and not on the location of the hosts. In this work we propose an enhanced routing protocol of our previous topology-oblivious \textbf{M}ultihop, \textbf{M}ultipath and \textbf{M}ultichannel NDN for VANETs (MMM-VNDN) routing strategy that exploits several paths to achieve more efficient content retrieval. Our new enhanced protocol, \textbf{i}mproved MMM-VNDN (iMMM-VNDN), creates paths between a requester node and a provider by broadcasting Interest messages. When a provider responds with a Data message to a broadcast Interest message, we create unicast routes between nodes, by using the MAC address(es) as the distinct address(es) of each node. iMMM-VNDN extracts and thus creates routes based on the MAC addresses from the strategy layer of an NDN node. Simulation results show that our routing strategy performs better than other state of the art strategies in terms of Interest Satisfaction Rate, while keeping the latency and jitter of messages low.
\end{abstract}

\section{Introduction}
\label{intro}
\input{intro.tex}

\section{Background and Related Work}
\input{related.tex}
\label{related}

\section{iMMM-VNDN: Description of main concepts}
\input{architecture.tex}
\label{architectures}

\section{Performance Evaluation}
\input{results.tex}
\label{evaluation}

\section{Conclusions}
\input{conclusion.tex}
\label{conclusion}

\section*{Acknowledgments}
This work was undertaken under the CONTACT project, CORE/SWISS/15/IS/10487418,
funded by the National Research Fund Luxembourg (FNR) and the Swiss National Science
Foundation (SNSF).

\bibliographystyle{unsrt}
\bibliography{bibliography.bbl}

\end{document}

%% file: intro.tex
Vehicular Ad Hoc Networks (VANETs) are a sub-category of MANETs~\cite{giordano2002mobile}. 
VANET applications are divided mostly into two categories: infotainment and safety applications~\cite{hartenstein2008tutorial}. Infotainment applications include video streaming, navigation and advertisements, while safety applications include traffic information, road accident warnings and/or weather conditions. These applications require a stable end-to-end connection with a source like a server, which holds some information (e.g., a highly requested video). This information source may be located far away from the requester node.  
Since VANETs consider vehicles, end-to-end connections are not stable, because the location of a vehicle could change in an unpredictable way while traveling with high speed. In addition, even if an end-to-end connection has been established, varying wireless channel conditions may lead to signal loss and/or bad Quality of Service (QoS) and Quality of Experience (QoE) (e.g., freezing of a video, delay in a safety notification). 

To address these issues, Named Data Networking (NDN) in VANETs has been proposed~\cite{chen2014vendnet}. NDN is a variant of Information Centric Networking (ICN), where information is forwarded based on content names but not on the location of the hosts~\cite{zhang2014named}. NDN messages are described with a unique name. Routing and forwarding decisions are based on this name. NDN mechanisms are described in Section~\ref{ndn}.

This paper presents an enhanced routing protocol of our previous work~\cite{kalogeiton2017multihop}, which presents a Vehicle to Vehicle (V2V) Multihop and Multipath routing protocol for NDN-VANETs. The main drawback of \cite{kalogeiton2017multihop} is that it always uses broadcast MAC addresses to flood messages in the network. Hence, to perform the forwarding decisions of an NDN message new fields in the NDN messages are created. The new protocol does not create additional fields in the NDN messages, therefore, the message overhead is lower than in \cite{kalogeiton2017multihop}. 
Instead, the proposed protocol floods Interest messages with broadcast MAC addresses, and when the content source responds with a Data message, we create unicast routes in vehicles that target destination nodes. 
Moreover, our new protocol uses IEEE 802.11p that has been proposed for VANETs \cite{jiang2008ieee} as the communication standard on top of NDN, instead of IEEE 802.11a that is used in \cite{kalogeiton2017multihop}. 
Furthermore, in this work we present extended experimental evaluations based on real traffic information from the Luxembourg map~\cite{codeca2015luxembourg}.

 We developed an enhanced protocol of our \textit{topology-oblivious V2V \textbf{M}ultihop, \textbf{M}ultipath and \textbf{M}ultichannel} routing strategy for \textbf{V}ANETs using \textbf{NDN}, MMM-VNDN \cite{kalogeiton2017multihop}. In this work, we present \textit{\textbf{i}mproved} MMM-VNDN, \textbf{iMMM-VNDN}, which creates unicast paths from a requester to a content source. iMMM-VNDN has the following characteri-stics:  
\begin{itemize}
\item We flood Interest messages with broadcast MAC addresses to discover potential content sources. Thus, all nodes receiving such a message, will continue flooding it.
\item When a content source responds with content to \textit{one} broadcast Interest message, we create  hop-by-hop routes by using MAC unicast addresses in each node. The combination of these routes leads to paths between the requester node and the source node. 
\item We exploit these routes to target next nodes, by defining destination MAC addresses to forward the next Interest messages by unicast.
\item For each route we measure the latency of the route and how many times this route was selected for forwarding. With this information we develop three different approaches for route selection: 
\begin{enumerate}[label=(\roman*), wide, labelindent=0pt, labelwidth=!]
\item We distribute the traffic uniformly to all recently created available routes.
\item We choose a route based on the lowest latency of the route.
\item We combine the two previous path selection approaches by distributing traffic uniformly to all recently created available routes with the lowest latency.
\end{enumerate}
\end{itemize}
We use the MAC addresses of each node as a main identifier to identify specific nodes. Thus, we create unicast transmissions by addressing each node by its MAC address(es). 
The main difference between the two protocols, MMM-VNDN \cite{kalogeiton2017multihop} and the proposed iMMM-VNDN is on the NDN message transmission.
In MMM-VNDN we created two \textbf{new} fields in the NDN messages, called Target MAC Address (TMA), which contains the MAC address of the next hop (target node) and Origin MAC Address (OMA), which contains the MAC address of the node that forwards the message. In MMM-VNDN we insert the MAC address of the interface of a node into the OMA (in either an Interest or a Data message) and the MAC address of the next node that should receive the message into the TMA. Then, we always use broadcast MAC addresses to flood this message into the network, and based on the TMA that is extracted from the NDN messages, we accept or reject incoming messages. Specifically, the OMA and TMA decide whether a node will accept or reject a message on top of the broadcast transmission. 

In iMMM-VNDN, we extract the OMA and the TMA from the strategy layer of the node, and we use them as fields inside the strategy layer. Thus, we leave the original NDN messages \textit{unchanged}. The strategy layer of NDN is equivalent to the data link layer of the OSI model. Then, we create unicast transmissions to send messages by using the OMA as source address and the TMA as destination addresses. 
We show that iMMM-VNDN allows the network to support high mobility and speeds of vehicles, and enables the VANET to adjust to many scenarios that could happen, e.g., content source moves out of range of requester etc. In addition, by reducing the broadcast transmissions of messages (since we create unicast transmissions) and by containing no additional information in the messages iMMM-VNDN achieves better results, in terms of content retrieval and latency, compared to MMM-VNDN and other state of the art protocols. In iMMM-VNDN a message is discarded in the strategy layer of the NDN stack, when the unicast MAC address of the message is not the same as the MAC address of the node, thus, allowing us to suppress duplicate transmissions and have a local overview of the incoming messages. 
The main characteristic is that in iMMM-VNDN we do not attach additional information to the NDN messages, thus, the overhead of NDN messages remains the same. Moreover, by making the decision in one of the lower levels of the NDN stack, we allow the upper layers to make decisions about only forwarding a message.

The rest of this paper is organized as follows: Section 2 presents an overview of related work about VANETs and NDN. We introduce our architecture and present the developed forwarding decisions in Section 3. We present the performance evaluation in Section 4. Finally, we draw our conclusions in Section~\ref{conclusion}.  

%% file: related.tex
\subsection{Routing in VANETs}
Authors in \cite{abedi2008enhancing} propose an extension of AODV, an on-demand reacting routing protocol \cite{perkins2003ad}, where the direction of the vehicle is the main parameter to determine the next hop. Moreover, in~\cite{toutouh2012intelligent} OLSR~\cite{clausen2003optimized} is extended, which is a proactive link state routing algorithm, by using an automatic optimization tool to define optimal parameters. Cluster Based Routing (CBR)~\cite{luo2010new} is a routing protocol that divides the geographical area into grids and assigns a vehicle as a cluster head to each grid to forward the packet to the destination node. 
VNIBR~\cite{saians2017efficient} is a routing protocol that is running on top of a virtualization layer in VANETs. This layer divides the geographical area into regions (clusters) and in each region there are three types of virtual nodes, each responsible for handling packets differently.~\cite{abdou2015priority} deals with the broadcasting problem in VANETs and proposes an Autonomic Dissemination Method (ADM), a method that delivers messages according to the network density and the given priority level of a message. 
\subsection{Named Data Networking} \label{ndn}
Named Data Networking (NDN)~\cite{zhang2014named} has been introduced to solve some of the problems that arise with today's TCP/IP host-to-host based network model \cite{arjunwadkar2014introduction}.
There are two NDN packet types, Interest and Data packets. A requester node sends an Interest (request for some information) and waits for the Data packet to be returned from a content source, thus satisfying the Interest.
Every node has it's own data structures: the Pending Interest Table (PIT), the Content Store (CS), and the Forwarding Information Base (FIB). The PIT stores information about already forwarded, but not yet satisfied, Interests. The CS stores previously received Data packets and the FIB stores information on how to forward Interests, based on their name. 
When an Interest arrives at a node, the CS is checked to determine whether a previously cached Data packet can satisfy the Interest. If no Data can be found in the CS, the PIT entry is checked. An existing PIT entry means that the Data has been requested already and further forwarding of the Interest is not necessary. 
If there is no PIT entry, the FIB is checked, a PIT entry is created, and the Interest is forwarded upstream towards the content source. Once the content source has been reached, the Data is sent downstream following the PIT entries of the Interest.   
\subsection{NDN in VANETs}\label{ndnvanets}
NDN has been proposed for V2V (Vehicle to Vehicle), V2I (Vehicle to Infrastructure) and V2R (Vehicle to Roadside Unit Communication)~\cite{chen2014vendnet}. One of the problems is that NDN does not allow forwarding a packet through its incoming face, which is sound for wired networks but poses unnecessary restrictions on Wi-Fi networks, especially with low numbers of network interfaces per node~\cite{amadeo2015forwarding}. The authors solved the multi-hop wireless issue by allowing rebroadcasting of an Interest through its incoming Wi-Fi face. In this work we add multiple network interfaces in one node and we also allow the rebroadcasting of an Interest through its incoming Wi-Fi face.

In~\cite{grassi2014vanet} the addition of GPS location services in the NDN-VANET implementation is proposed, to get the geographical distance between entities (vehicles, infrastructure, etc.) and let the strategy decide which node is farthest away from the sender for continued propagation. 
Also in~\cite{grassi2014vanet} authors implemented an opportunistic caching strategy for the CS. In~\cite{anastasiades2016dynamic} Interests are broadcast to identify paths between requesters and content sources, and a new data structure is created, called Content Request Tracker (CRT), to decide whether an Interest should be broadcast (in case there are many requesters) or be sent along a unicast path. In~\cite{Amadeo2013enhancing,amadeo2012contentcentric,amadeo2012upcomingVanets} Content-Centric Vehicular Networking (CCVN) for general-purpose multi-hop content distribution is introduced. CCVN is based on the CCN ideas~\cite{jacobson2009networking} and is compliant with the Wireless Access in Vehicular Environments (WAVE) architecture~\cite{893287}. A controlled Data and Interest propagation sche-me (CODIE) is presented in \cite{ahmed2016codie}, where each node and each message maintains a hop count indicating the minimum hop count to reach a destination. If the hop count is exceeded, the message is discarded. This approach reduces the flooding of a Data packet, by controlling its dissemination in the network. The network though is still burdened with all Data packets that exist in it.
In~\cite{gomes2017addressing} authors develop two solutions for low densities of vehicles in a VANET: delegating content retrieval through infrastructure or taking advantage of the store, carry, and forward mechanism by vehicles and retransmitting NDN messages from these vehicles. \cite{ncc} presents NCC, a forwarding strategy that sends an Interest through the face with the lowest prediction time. In~\cite{kalogeiton2017sdn} the combination of NDN with Software Defined Networking (SDN) is proposed and a network architecture in VANETs is presented. \cite{duarte2017multi} combines NDN, SDN, and Floating Content to efficiently disseminate messages in VANETs. 

Our work does not rely on geographical information, since we cannot always assume that we know the location of each node. Therefore, we use omni-directional antennas in the network nodes, to cover all geographical locations around a node (360\degree), in order to reach every vehicle around a node. Moreover, this study does not require any infrastructure assistance, since until today many cities do not support infrastructure for VANETs (e.g., Road Side Units - RSUs). Finally, we propose to use the existing NDN data structures and NDN messages, and not to create new data structures to retrieve content, allowing us to be compatible with other NDN devices (such as servers that support the default NDN data structures). 

%% file: architecture.tex
\subsection{Overview}\label{overview}
A main characteristic of VANETs is the mobility of nodes. Thus, traditional TCP/IP fails, when an end-to-end path between requester and source breaks before content retrieval. Cellular network infrastructure has also been proposed for VANETs. \cite{mir2014lte} proposes LTE in VANETs because of its high scalability and its mobility support. However, ad-hoc networks cannot rely only on cellular infrastructure for all communications. 
We emphasize that the use of V2V communication is crucial in a VANET, since it allows message exchange and direct communication among vehicles in a restricted local area, hence reducing network delays. Moreover, cellular network infrastructure could be used for VANETs as a backup mechanism, e.g., for content retrieval, when the content is unavailable in an area, or the content does not exist in local caches of the vehicles anymore.  

Traditional MANET routing protocols require either neighbor discovery by broadcasting HELLO messages (e.g. AODV \cite{perkins2003ad}) or knowledge of the content location, such as GPSR \cite{karp2000gpsr}. However, in VANETs we can not assume that the location of the content will be known to everyone in the network, or that the content will always be kept at the same node or location. In our proposed routing protocol, we flood Interest messages with a broadcast MAC address to discover potential content sources, and to create paths. Thus, we avoid broadcasting additional messages, such as HELLO messages, that would burden the network. 

The new routing protocol is presented in this Section. 
iMMM-VNDN discovers content sources, creates paths, and forwards messages based on information that these paths provide.
Our algorithm saves network resources by minimizing the possible transmissions in nodes. The main characteristic is the creation of paths, by creating unicast routes from broadcast transmissions (as in Wi-Fi communications). The advantage of path creation is that a limited number of nodes participate in the content retrieval process, thus saving network resources, which are available for other applications. To avoid redundant transmissions and to target destination nodes we add a unique identifier to the PIT and the FIB table in each node: the MAC address(es) of a node's interface(s). We explain the reason behind this approach, together with the forwarding of Data and Interest messages, in Subsection~\ref{routing}.  
\subsection{Routing}\label{routing}
In both of our algorithms, i.e. \textbf{M}ultihop, \textbf{M}ultipath and \textbf{M}ulti-channel for \textbf{V}ANETs using \textbf{NDN} (\textbf{MMM-VNDN}) and \textit{\textbf{i}mpro-ved} \textbf{MMM-VNDN} (\textbf{iMMM-VNDN}), the first task is to identify the content source(s) and to create paths between it and the requester node(s), which can be used to receive messages.  

iMMM-VNDN creates unicast transmissions in VANETs, by using a unique identifier to distinguish the nodes. We use the MAC address(es) of a node's interface as unique node identifiers.
Thus, the current NDN implementation has been extended and developed by including the following: 

\begin{itemize}
    \item Target MAC Address (TMA) is the destination MAC address of the message. Thus, it shows the next hop that the Interest or Data message will be forwarded to. 
    \item Origin MAC Address (OMA) is the source MAC address of the message. It shows the network device of the node that the Interest or Data message has been forwarded from (previous hop). 
\end{itemize}
OMA and TMA assist in identifying intermediate nodes forwarding messages and creating paths, cf. Section~\ref{second}. In addition, new fields in the PIT and FIB tables of every node have been added to the existing NDN implementation.

In MMM-VNDN we create additional fields into the NDN messages that include the OMA and the TMA. Instead, in iMMM-VNDN we leave the NDN header unchanged. Therefore, we do not use the OMA and the TMA as fields in the NDN message. We extract the MAC addresses, the origin MAC address and the target MAC address, from the strategy layer at each NDN node. Then, we use the OMA and the TMA as fields inside the strategy layer of each node to perform routing decisions. 
\begin{figure}[t]%
    \centering
    \subfloat[Broadcasting of the Interest by the requester node for the first time - Flooding]{{\includegraphics[width=0.87\linewidth] {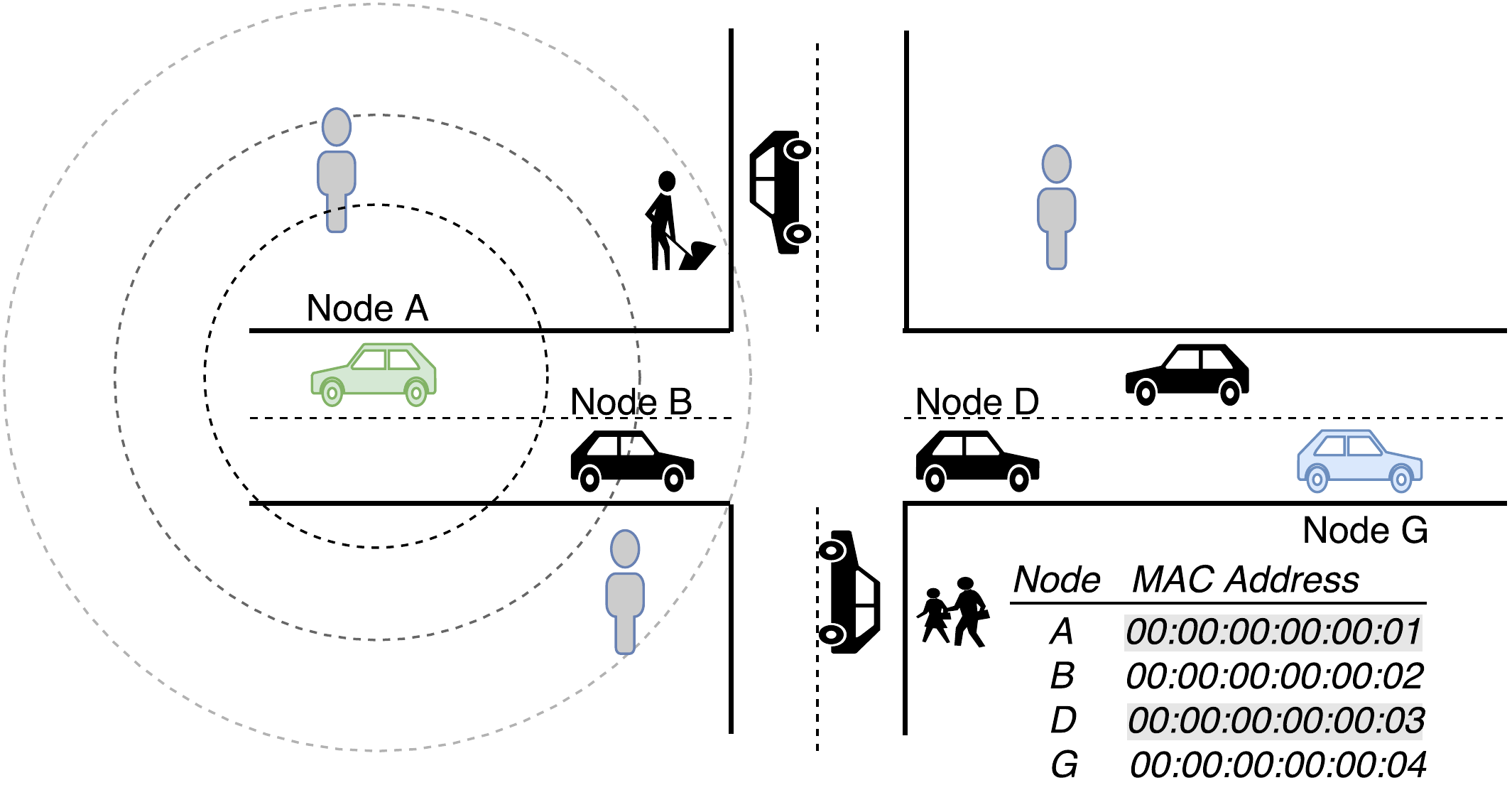} }
    \label{broadcasteirini}}

    \subfloat[Interest processing and forwarding from node B]{{\includegraphics[width=0.75\linewidth]{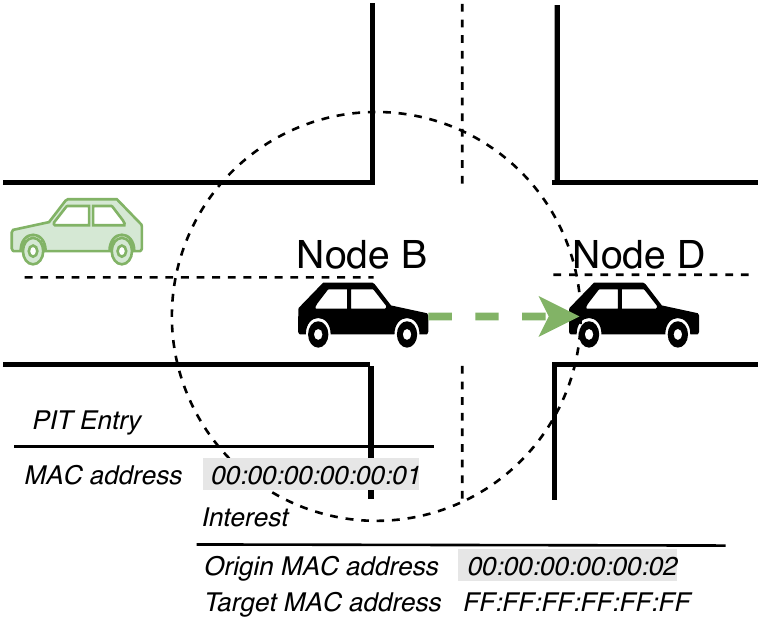} }  
    \label{1pathc} }
    \caption{Flooding phase}%
    \label{fig:example}%
\end{figure}
\subsubsection{First phase - Flooding}\label{flooding}
To begin with, every node has an empty FIB table. Thus, it should populate its FIB table to identify possible next hops towards the content source.  
A node that requests content (requester), first floods an Interest message to request a segment of the content. This Interest message contains in its OMA (source address) the MAC address of the interface of the node that has forwarded the Interest message (in this case the requester). The TMA in this Interest message is a broadcast MAC address.

\begin{figure}%
    \centering
    \subfloat[Node G responding with Data]{{\includegraphics[width=0.47\linewidth]{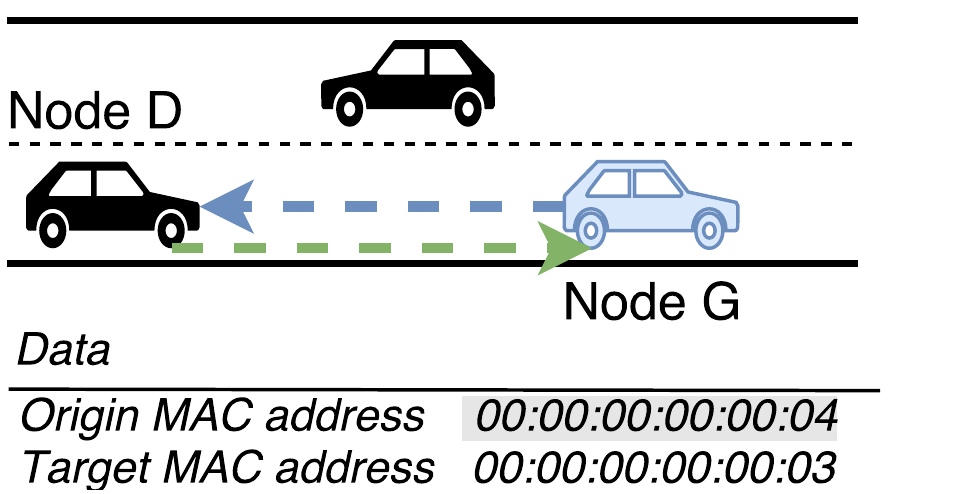} }\label{2a}}%
      \subfloat[Node D processes and forwards the Data]{{\includegraphics[width=0.47\linewidth]{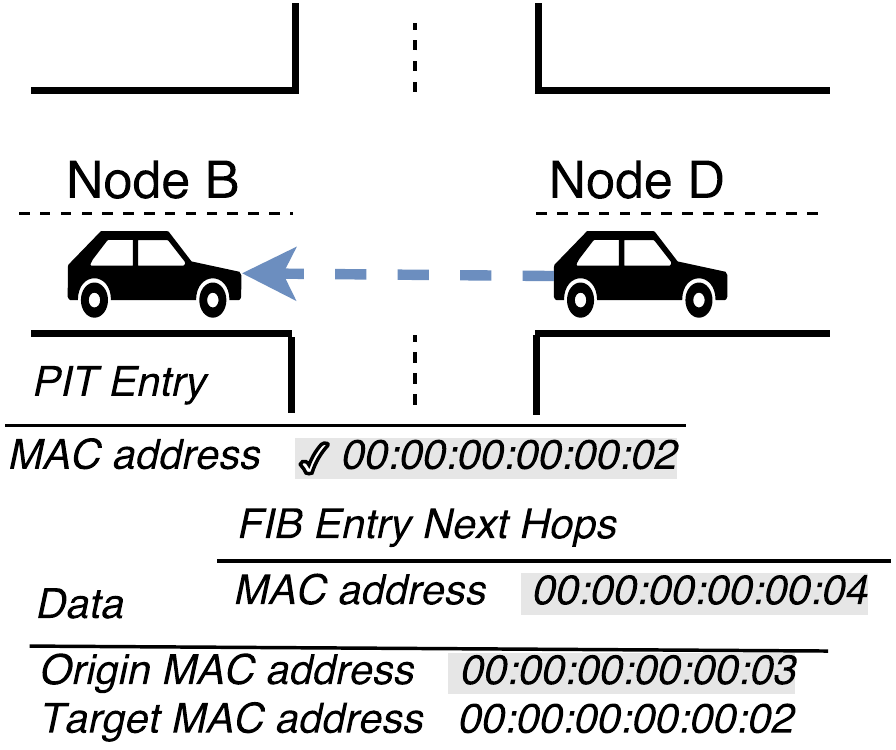} }\label{2b}}%
    \caption{Data processing}%
    \label{fig:example}%
\end{figure}

Let us assume that the topology of a VANET is as shown in Fig.~\ref{broadcasteirini}. The MAC addresses of the participating nodes are also shown in Fig.~\ref{broadcasteirini}. Node A is the requester node and node G is the content source. First, node A floods an Interest containing 00:00:00:00:00:01 as OMA and FF:FF:FF:FF:FF:FF as TMA, which is the MAC broadcast address. When an intermediate node receives this Interest message, it checks the OMA to identify the node that sent the message. Then, this node checks its CS, to identify if it has the requested Data already cached. If the CS of the node does not contain the requested Data, the node creates or updates a PIT entry, containing the OMA. Since the TMA is broadcast, this intermediate node continues flooding the message with the broadcast MAC address. As shown in Fig.~\ref{1pathc}, node B that received the broadcast Interest from node A, creates a PIT entry with MAC address: 00:00:00:00:00:01. Node B updates the Interest's OMA to 00:00:00:00:00:02 and floods the message with TMA: FF:FF:FF:FF:FF:FF. This process continues until this Interest message arrives at the node holding the requested Data. In iMMM-VNDN, the content source, after receiving the Interest, responds with a Data message and performs the following actions: 
\begin{enumerate}[label=(\roman*)]
    \item It takes the OMA (source address of the previous hop) of the Interest message and inserts it into the Data message as its TMA (destination MAC address). 
    \item It takes its own MAC address and inserts it into the OMA of the Data message.
    \item It unicasts the Data message into the network, to \textbf{all} nodes that previously broadcast the Interest message to it. 
\end{enumerate}
This process is illustrated in Fig.~\ref{2a}. 
Node G, which is the content source, receives an Interest that has a broadcast TMA and OMA: 00:00:00:00:00:03. Node G, then, creates a Data message by entering 00:00:00:00:00:03 into the TMA and 00:00:00:00:00:04 into the OMA. Then, it unicasts the Data message into the network.

When an intermediate node receives a Data message it checks if the Data message is meant for itself, i.e. if the TMA (destination address) in the Data message is the same as its own MAC address. If not, the message is discarded. If the TMA of the message and the node's MAC address are the same, the node checks the PIT table. If there is no matching PIT entry, the Data message is discarded. If there exists a PIT entry, the node performs the following actions:
\begin{enumerate}[label=(\roman*)]
    \item The node creates (or updates), a new FIB entry that contains the OMA of the Data message (source address of the previous hop of the message).
    \item The node updates the OMA of the Data message to contain its own MAC address.
    \item The node checks the PIT entry that has been created from the respective Interest, and sets the TMA of the Data message to the MAC address of the PIT entry. 
    \item The node enters the Data packet in its CS. 
    \item The node sends via unicast the updated Data message to the network.
\end{enumerate}
\begin{algorithm}
\caption{Data forwarding from intermediate node}
\label{Data Forwarding}
\begin{algorithmic}[1]
\Procedure{Reception of Data}{}
\If{$PIT Entry \neq \emptyset$}
    \If{ $TargetMAC\neq MyMAC$} 
    \State $Create/Update(FIBEntry,OriginMac)$ 
    \Else 
    \State $Create/Update(FIBEntry,OriginMac)$ 
    \State $OriginMAC\gets MyMAC$
    \State $TargetMAC\gets PITEntry(nexthop)$
    \State $forward(Data,nexthop)$
    \EndIf
\EndIf
\EndProcedure
\end{algorithmic}
\end{algorithm}
The above process is outlined in Algorithm~\ref{Data Forwarding}. To continue with our example, the Data message sent by node G in Fig.~\ref{2a} is now coming in at node D. As shown in Fig.~\ref{2b} node D first checks the TMA that is set to 00:00:00:00:00:03 (as shown in Fig.~\ref{2a}). Since it is the same as node D's MAC address, the message is accepted. Then, node D checks, if a PIT entry exists for this Data message. Since a PIT entry exists, node D creates a FIB entry with 00:00:00:00:00:04 and updates the TMA to 00:00:00:00:00:02, and the OMA to 00:00:00:00:00:03.

This process continues until the requester receives the Data message.
All nodes will continue flooding messages, until they have one FIB entry in their FIB table. 
This means that a node will continue the flooding of Interests with broadcast MAC address, until a route to a next hop is available.
\begin{figure}[t]%
	\centering
	\subfloat[Interest forwarding from node A]{{\includegraphics[width=0.85\linewidth]{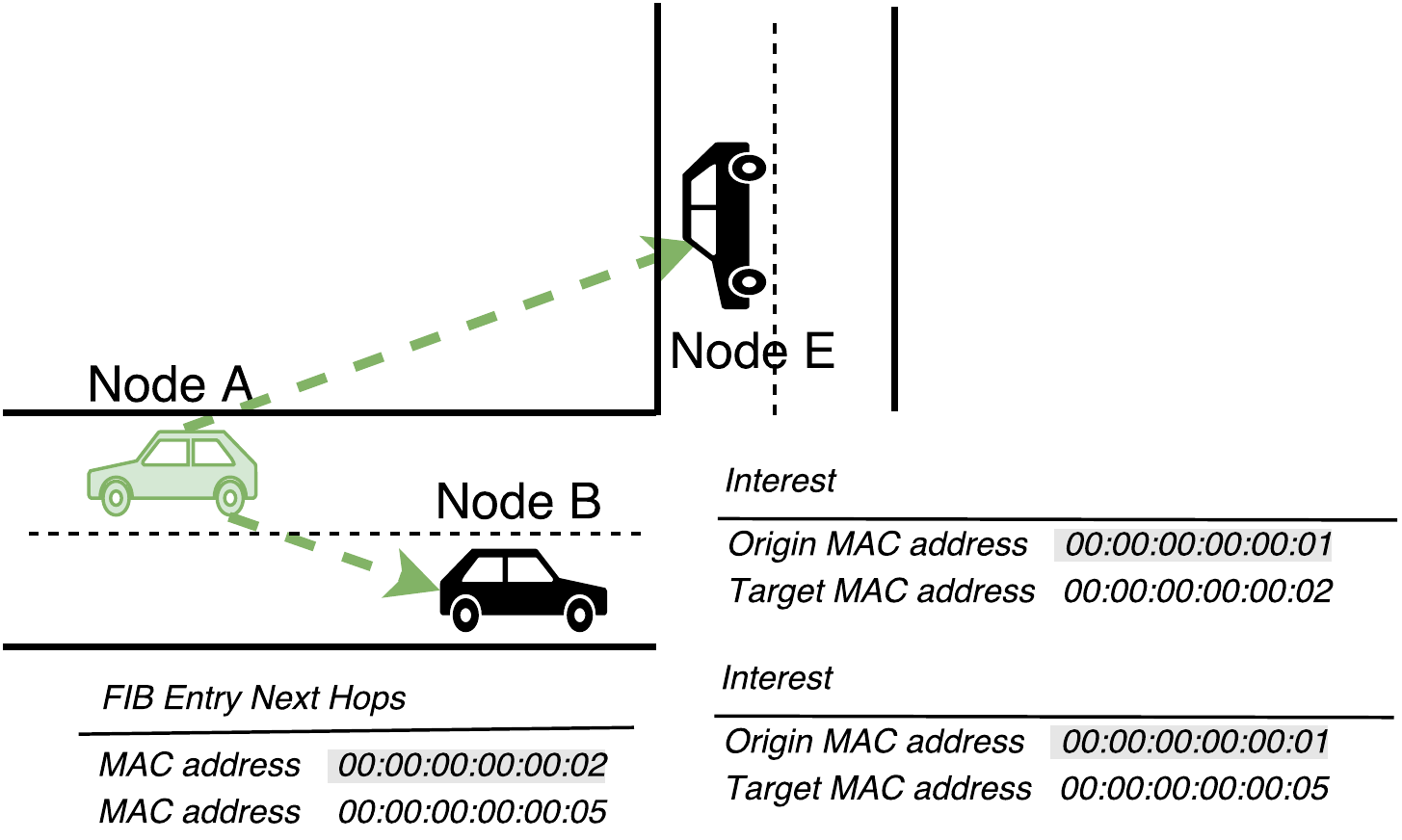} }  
		\label{3b}}
	\\
	\centering
	\subfloat[Interest processing and forwarding from node E]{{\includegraphics[width=0.85\linewidth]{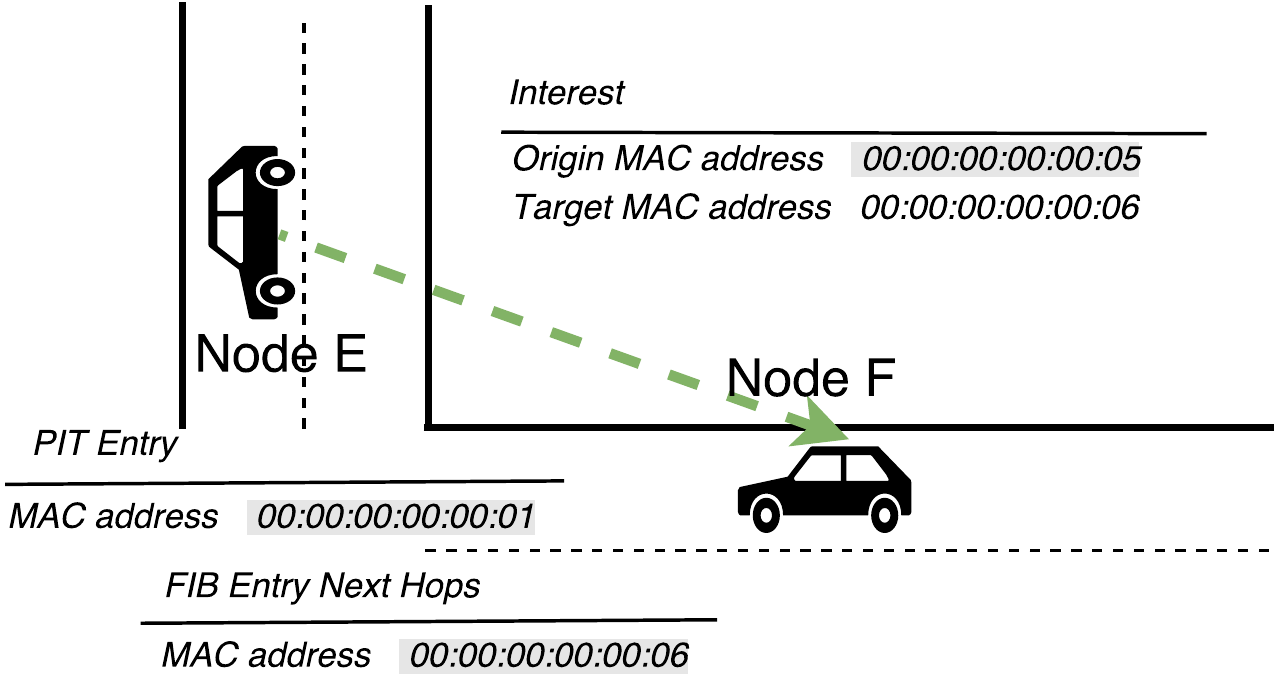} }  
		\label{3c} }
	\caption{Forwarding based on FIB entries}%
	\label{fig:example2}%
\end{figure}
\subsubsection{Second phase - Forwarding based on FIB}\label{second}
After the first phase (cf. Section~\ref{flooding}) the requester(s) and intermediate nodes will have received the same Data message by different nodes, and will have created FIB entries that will contain multiple MAC addresses (multiple next hops).  
If several next hops exist, a node should select one of them. Hence, the FIB was extended with \textit{two} additional fields, as shown in Table~\ref{nexthop}. For the selection of the next hop of the Interest, we propose three \textbf{path selection approaches}. 
\begin{enumerate}[label=(\roman*), wide, labelindent=0pt, labelwidth=!]
\item When a FIB next hop is created, the counter field is set to 0. Every time this next hop is chosen for forwarding, the counter field is incremented by one. The first approach is selecting the next hop field that corresponds to the lowest counter. This ensures that we distribute the traffic to all possible next hops. When there are multiple next hops with the same counter, the selection is based on the last added next hop of the FIB table. 
\item The second approach is to choose the next hop with the lowest latency. Latency is defined as the time that has passed from the transmission of the Interest message to the reception of the Data message in a specific node. This latency is included in the respective FIB entry. If the path is chosen several times, the latency field resets and shows always the last counted latency. Based on Table~\ref{nexthop}, node A chooses the second next hop, because of the lowest latency, with a MAC address of 00:00:00:00:00:05. 
\begin{table}[t!]
	\caption{FIB Entry next hop additional fields}\label{nexthop}
	\begin{center}
		\begin{tabular}{l*{6}{c}r}
			MAC address    & Latency(ms) &  Counter \\
			\hline
			00:00:00:00:00:02      & 100    & 0      \\  
			00:00:00:00:00:05      & 50     & 0     \\
			\lasthline
		\end{tabular}
	\end{center}
\end{table}
\item The third approach is based on the combination of the two approaches above. Each time when a next hop is selected, the counter is increased by one. When multiple next hops exist, which have been added almost at the same time and have not been selected for forwarding, their counters have the same value. In that case, the next hop with the lowest latency is chosen. 
For the values that are shown in Table~\ref{nexthop}, their counter field is zero. Hence, the latency field will be checked, and the next hop with the lowest latency will be selected, i.e. 00:00:00:00:00:05. 

Subsequently, when the requester node sends the second Interest it checks the FIB table to identify possible routes. It searches the FIB for possible MAC addresses of next hops, i.e. destination MAC addresses of next nodes to unicast the Interest.
\begin{table*}[h]\centering
	\caption{Simulation Parameters}\label{table2}
	\begin{center}
		\begin{tabular}{| >{\centering\arraybackslash}m{1.8cm}|| >{\centering\arraybackslash}m{2.8cm}||>{\centering\arraybackslash}m{2cm}||>{\centering\arraybackslash}m{1.3cm} || >{\centering\arraybackslash}m{1.9cm}|}
			\hline
			\centering{\arraybackslash}{\textsc{\textbf{Strategy}}} &\centering{\textsc{\textbf{ Propagation Loss Model}}} & \centering{\textsc{\textbf{Standard-Transmission Range}}} & \centering{\textsc{\textbf{Data Rate}}}  & \centering{\textsc{\textbf{Channel Bandwidth}}}
			\tabularnewline
			\hline   
			\centering MMM-VNDN  &  \centering Three Log Distance \newline and Nakagami  &\centering IEEE802.11a  \newline200m  &  24 Mbps &20MHz \tabularnewline  
			\hline
			\centering iMMM-VNDN & \centering Two Ray Ground   &\centering IEEE802.11p  \newline250m  &6 Mbps  & 10MHz\tabularnewline 
			\hline
			\centering Flooding & \centering Three Log Distance \newline and Nakagami &\centering IEEE802.11a \newline 200m  &  24 Mbps&20MHz \tabularnewline
			\hline
			\centering Best-route & \centering Three Log Distance \newline and Nakagami  &\centering IEEE802.11a  \newline200m&  24 Mbps&20MHz \tabularnewline
			\hline
			\centering NCC & \centering Three Log Distance \newline and Nakagami  &\centering IEEE802.11a  \newline200m &  24 Mbps &20MHz    \tabularnewline
			\hline
			\centering CCVN& \centering Two Ray Ground  &\centering IEEE802.11p  \newline250m  &6 Mbps & 10MHz \tabularnewline
			\hline
			\centering CODIE & \centering Two Ray Ground    &\centering IEEE802.11p  \newline250m &6 Mbps & 10MHz  \tabularnewline \hline
		\end{tabular}
	\end{center}
\end{table*}
\end{enumerate}
In iMMM-VNDN the requester selects a next hop and sets the Interest fields as follows: The OMA contains the current node's MAC address and the TMA contains the MAC address of the next hop that has been selected as the destination MAC address from its FIB. Node A, as shown in Fig.~\ref{3b}, checks its FIB, where it identifies two possible next hops: 00:00:00:00:00:02 and 00:00:00:00:00:05. It selects one (according to the above path selection approaches), updates the TMA of the Interest to 00:00:00:00:00:05 and the OMA to 00:00:00:00:00:01. When a node receives an Interest containing a TMA, it checks if the latter is the same MAC address as its own. If the TMA is different, it discards the message. If the TMA is the same as its own, the node accepts the Interest. Then, the node checks if there is a match in its CS. If no match is found, the node enters the OMA from the Interest message to the PIT. Then, it checks its FIB table in order to identify possible next hops. After, it forwards the message with an updated OMA that is set to its own MAC address and a TMA that is set to the MAC address of the chosen next hop. 
Fig.~\ref{3c} shows this process, where an incoming Interest arrives at node E. Node E checks the TMA of the Interest: 00:00:00:00:00:05, creates a PIT entry with 00:00:00:00:00:01, and searches its FIB for possible next hops: 00:00:00:00:00:06. Then it updates the OMA of the Interest to 
00:00:00:00:00:05 and the TMA of the Interest to 00:00:00:00:00:06.
This process continues until the Interest message arrives at the content source. The responding Data message follows the reverse path, as described in Section~\ref{flooding}.

\subsection{Node discovery and  FIB deletion}\label{nodediscovery}
Since the topology in VANETs is constantly changing, created paths from a requester to a content source may break unexpectedly.
In order to discover new routes and new content sources, the chosen approach is to flood an Interest eve-ry few seconds with a broadcast MAC address. The FIB is populated regularly with new and active connections. When new vehicles in the network are discovered, they are also used for forwarding, and new routes that include them are created. Every time when an Interest message is flooded with a broadcast MAC address, the FIB table entries are deleted. In this way we control the FIB size and update it accordingly every time a new route is created. 

%% file: results.tex
\subsection{Path Selection Approaches and Metrics}\label{scenarios}
To evaluate the routing protocols, we used the three different path selection approaches as described in Section~\ref{second}.
 To evaluate the routing protocols three metrics were used: 
   \begin{itemize}
     	\item Interest Satisfaction Rate (ISR) describes the number of received Data messages that were received by the requester divided by the total number of Interest messages being sent.
     	\item Average Latency: The average latency for all received Data messages describes the average time that passed from the time a requester sent an Interest message to the time that the requester received the Data message. When a Data has not been received until its expiration time, the requester node retransmits the Interest message and the initial time is reset.
     	\item Average Jitter: Jitter is defined as the mean deviation of the difference in packet spacing at the receiver node compared to the sender for a pair of packets~\cite{rfc1889}. 
     \end{itemize}
     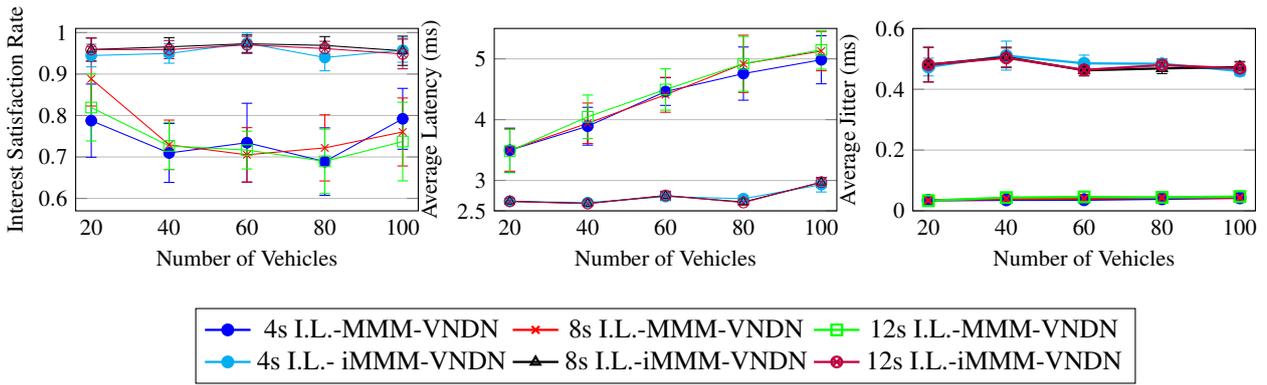
\begin{figure*}[!ht]
        	\centering
	\begin{tikzpicture}
    \begin{groupplot}[legend columns=3,legend to name= Fig7Legend,
            legend entries={{\normalsize 4s I.L.-MMM-VNDN},{\normalsize 8s I.L.-MMM-VNDN},{\normalsize 12s I.L.-MMM-VNDN},{\normalsize 4s I.L.- iMMM-VNDN},{\normalsize 8s I.L.-iMMM-VNDN},{\normalsize 12s I.L.-iMMM-VNDN}},
            group style={group size= 3 by 1, group name = fig7_plots},
	width  =0.35\textwidth,
	height = 4cm,
	major x tick style = transparent,
	ymajorgrids = true,
	xlabel = {Number of Vehicles},
	symbolic x coords={20,40,60,80,100},
	xtick = data,
	scaled y ticks = false,
	enlarge x limits=0.05,
	ymin=0.57,
	ymax=1.01,
	ylabel near ticks]
        \nextgroupplot[ylabel={Interest Satisfaction Rate }]
    \addplot[
        mark=*,
        blue,
        error bars/.cd, y dir=both, y explicit,
    ] plot coordinates {
    	(20, 0.7876)+= (0,0.0883) -= (0,0.0883)	
    	(40, 0.7096)+= (0, 0.0711) -= (0, 0.0711)    	
	    (60, 0.734376503) +=(0, 0.09517738) -= (0, 0.09517738)		
        (80, 0.6888141) +=(0, 0.081398654) -= (0,0.081398654)
        (100, 0.792010201)+=(0, 0.073466565) -= (0, 0.073466565)
    };
    \addplot[
    color=red,mark=x,error bars/.cd, y dir=both, y explicit
    ]
        plot coordinates {
        	(20, 0.8884) += (0, 0.0656) -= (0,  0.0656)
     		(40, 0.7293) += (0, 0.0597) -= (0, 0.0597)
     	 (60, 0.705226009) += (0,0.065490618) -= (0,0.065490618)	
     	 (80, 0.721910058)+= (0, 0.079785695) -= (0, 0.079785695)
     	 (100, 0.760264271)+= (0,0.082013012) -= (0, 0.082013012)
        };
        \addplot[
        color=green,mark=square,error bars/.cd, y dir=both, y explicit
        ]
        plot coordinates {			
				(20, 0.8193)+= (0,0.0808) -= (0,0.0808)
     			(40, 0.7267)+= (0, 0.0569) -= (0,0.0569)
     		 (60, 0.71659009)+= (0,0.045641364) -= (0,0.045641364)	
     		 (80,	0.689402953)+= (0,0.077690666) -= (0,0.077690666)	
     		 (100, 0.737166207)+= (0,0.094847661) -= (0,0.094847661)	
        };
   \addplot[
   mark=*,
   cyan,
   error bars/.cd, y dir=both, y explicit,
   ] plot coordinates {
   (20, 0.944693) += (0,0.027328) -=(0,0.027328)
(40, 0.949685) += (0,0.023541) -=(0,0.023541)
(60, 0.974407) += (0,0.024529) -=(0,0.024529)
(80, 0.939927) += (0,0.032093) -=(0,0.032093)
(100, 0.957353) += (0,0.028792) -=(0,0.028792)
   };
   \addplot[
   color=black,mark=triangle,error bars/.cd, y dir=both, y explicit
   ]
   plot coordinates {		
   (20, 0.959166) += (0,0.027855) -=(0,0.027855)
(40, 0.965608) += (0,0.022282) -=(0,0.022282)
(60, 0.973100) += (0,0.021115) -=(0,0.021115)
(80, 0.969130) += (0,0.021302) -=(0,0.021302)
(100, 0.956114) += (0,0.035215) -=(0,0.035215)
   };
   \addplot[
   color=purple,mark=otimes,error bars/.cd, y dir=both, y explicit
   ]
   plot coordinates {			
   (20, 0.959166) += (0,0.027855) -=(0,0.027855)
(40, 0.959032) += (0,0.021786) -=(0,0.021786)
(60, 0.970428) += (0,0.020318) -=(0,0.020318)
(80, 0.961571) += (0,0.017319) -=(0,0.017319)
(100, 0.948385) += (0,0.035764) -=(0,0.035764)	
   };
    \nextgroupplot[ylabel={Average Latency (ms)}, ymin=2.5, ymax=5.5, ytick={2.5,3,4,5},scaled y ticks = false, log ticks with fixed point={1000 sep=},
	ylabel near ticks ]
    \addplot[
        mark=*,
        blue,
        error bars/.cd, y dir=both, y explicit,
    ] plot coordinates {
(20, 3.496263) += (0,0.363791) -=(0,0.363791)
(40, 3.892093) += (0,0.311368) -=(0,0.311368)
(60, 4.462207) += (0,0.227550) -=(0,0.227550)
(80, 4.757644) += (0,0.437208) -=(0,0.437208)
(100, 4.985125) += (0,0.395085) -=(0,0.395085)
    };
    \addplot[
    color=red,mark=x,error bars/.cd, y dir=both, y explicit
    ]
        plot coordinates {		
        	(20, 3.498761) += (0,0.346950) -=(0,0.346950)
(40, 3.940366) += (0,0.333422) -=(0,0.333422)
(60, 4.409115) += (0,0.287491) -=(0,0.287491)
(80, 4.918345) += (0,0.470175) -=(0,0.470175)
(100, 5.127839) += (0,0.323537) -=(0,0.323537)
        };
        \addplot[color=green,mark=square,error bars/.cd, y dir=both, y explicit]
        plot coordinates {		
     (20, 3.483540) += (0,0.355629) -=(0,0.355629)
(40, 4.046678) += (0,0.358090) -=(0,0.358090)
(60, 4.496185) += (0,0.340423) -=(0,0.340423)
(80, 4.918432) += (0,0.452082) -=(0,0.452082)
(100, 5.147093) += (0,0.316478) -=(0,0.316478)
        };  
     \addplot[
       mark=*,
     cyan,
     error bars/.cd, y dir=both, y explicit,
     ] plot coordinates {
     	(20, 2.655894) += (0,0.043643) -=(0,0.043643)
(40, 2.635812) += (0,0.032892) -=(0,0.032892)
(60, 2.734655) += (0,0.065073) -=(0,0.065073)
(80, 2.699700) += (0,0.054017) -=(0,0.054017)
(100, 2.930934) += (0,0.120794) -=(0,0.120794)
     };
     \addplot[
     color=black,mark=triangle,error bars/.cd, y dir=both, y explicit
     ]
     plot coordinates {		
     	(20, 2.656612) += (0,0.034991) -=(0,0.034991)
(40, 2.620867) += (0,0.029123) -=(0,0.029123)
(60, 2.747503) += (0,0.067443) -=(0,0.067443)
(80, 2.645608) += (0,0.039820) -=(0,0.039820)
(100, 2.968897) += (0,0.075220) -=(0,0.075220)
     };

     \addplot[
     color=purple,mark=otimes,error bars/.cd, y dir=both, y explicit
     ]
     plot coordinates {			
     	(20, 2.656612) += (0,0.034991) -=(0,0.034991)
(40, 2.619455) += (0,0.029371) -=(0,0.029371)
(60, 2.753552) += (0,0.069825) -=(0,0.069825)
(80, 2.636086) += (0,0.041552) -=(0,0.041552)
(100, 2.971773) += (0,0.074466) -=(0,0.074466)
     };
        \nextgroupplot[ylabel={Average Jitter (ms)}, ymin=0, ymax=0.6]
     \addplot[line width=0.30mm,
        mark=*,
        blue,
        error bars/.cd, y dir=both, y explicit,
    ] plot coordinates {
    	(20, 0.034696) += (0,0.008810) -=(0,0.008810)
(40, 0.036159) += (0,0.007550) -=(0,0.007550)
(60, 0.036563) += (0,0.006710) -=(0,0.006710)
(80, 0.039639) += (0,0.010751) -=(0,0.010751)
(100, 0.042651) += (0,0.009280) -=(0,0.009280)
    };
    \addplot[line width=0.30mm,
    color=red,mark=x,error bars/.cd, y dir=both, y explicit
    ]
        plot coordinates {		
        (20, 0.034608) += (0,0.011329) -=(0,0.011329)
(40, 0.039548) += (0,0.011399) -=(0,0.011399)
(60, 0.040301) += (0,0.013280) -=(0,0.013280)
(80, 0.043421) += (0,0.011863) -=(0,0.011863)
(100, 0.043242) += (0,0.012807) -=(0,0.012807)
        };
        \addplot[line width=0.30mm,
        color=green,mark=square,error bars/.cd, y dir=both, y explicit
        ]
        plot coordinates {			
       (20, 0.033416) += (0,0.007849) -=(0,0.007849)
(40, 0.043770) += (0,0.013819) -=(0,0.013819)
(60, 0.045842) += (0,0.017396) -=(0,0.017396)
(80, 0.044412) += (0,0.010903) -=(0,0.010903)
(100, 0.047103) += (0,0.015532) -=(0,0.015532)
        };
      \addplot[line width=0.30mm,
    mark=*,
    cyan,
    error bars/.cd, y dir=both, y explicit,
    ] plot coordinates {
    	(20, 0.473142) += (0,0.029209) -=(0,0.029209)
(40, 0.510612) += (0,0.047666) -=(0,0.047666)
(60, 0.485125) += (0,0.027203) -=(0,0.027203)
(80, 0.483857) += (0,0.013567) -=(0,0.013567)
(100, 0.459748) += (0,0.015944) -=(0,0.015944)
    };
    \addplot[line width=0.30mm,
    color=black,mark=triangle,error bars/.cd, y dir=both, y explicit
    ]
    plot coordinates {		
    	(20, 0.481054) += (0,0.057035) -=(0,0.057035)
(40, 0.505637) += (0,0.032633) -=(0,0.032633)
(60, 0.462266) += (0,0.018423) -=(0,0.018423)
(80, 0.468499) += (0,0.017091) -=(0,0.017091)
(100, 0.473075) += (0,0.018409) -=(0,0.018409)
    };
    \addplot[line width=0.30mm,
    color=purple,mark=otimes,error bars/.cd, y dir=both, y explicit
    ]
    plot coordinates {			
    	(20, 0.481054) += (0,0.057035) -=(0,0.057035)
(40, 0.503169) += (0,0.032366) -=(0,0.032366)
(60, 0.463663) += (0,0.017928) -=(0,0.017928)
(80, 0.479586) += (0,0.020127) -=(0,0.020127)
(100, 0.468651) += (0,0.020267) -=(0,0.020267)
    };
    \end{groupplot}
       legend style={
		at={(1,1.05)},
		anchor=south east,
		column sep=1ex}
     \node (fig7_Legend ) at ($(fig7_plots c2r1.center)-(0,3cm)$){\ref{Fig7Legend}};
       \end{tikzpicture}       
       \caption{Results in Manhattan map for the 1st proposed path selection approach}
    \label{scenario1isr}
    \end{figure*}

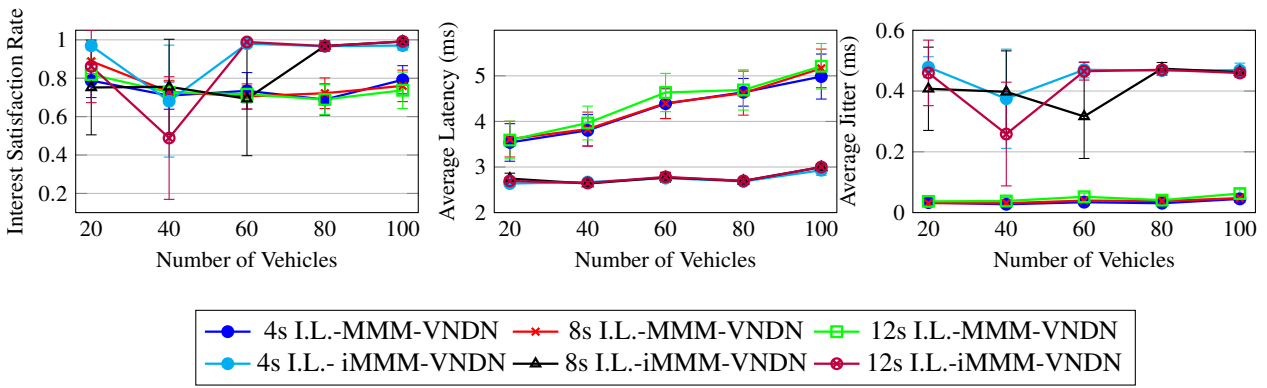
\begin{figure*}[!ht]
        	\centering
	\begin{tikzpicture}
    \begin{groupplot}[legend columns=3,legend to name= Fig6Legend,
            legend entries={{\normalsize 4s I.L.-MMM-VNDN},{\normalsize 8s I.L.-MMM-VNDN},{\normalsize 12s I.L.-MMM-VNDN},{\normalsize 4s I.L.- iMMM-VNDN},{\normalsize 8s I.L.-iMMM-VNDN},{\normalsize 12s I.L.-iMMM-VNDN}},
            group style={group size= 3 by 1, group name = fig6_plots},
	width  =0.35\textwidth,
	height = 4cm,
	major x tick style = transparent,
	ymajorgrids = true,
	ylabel = {Interest Satisfaction Rate},
	xlabel = {Number of Vehicles},
	symbolic x coords={20,40,60,80,100},
	xtick = data,
	scaled y ticks = false,
	enlarge x limits=0.05,
	ymin=0.1,
	ymax=1.05,
	ylabel near ticks]
        \nextgroupplot[ylabel={Interest Satisfaction Rate }]
    \addplot[line width=0.30mm,
               mark=*,
        blue,
        error bars/.cd, y dir=both, y explicit,
    ] plot coordinates {
    	(20, 0.7876)+= (0,0.0883) -= (0,0.0883)	
    	(40, 0.7096)+= (0, 0.0711) -= (0, 0.0711)    	
	    (60, 0.734376503) +=(0, 0.09517738) -= (0, 0.09517738)		
        (80, 0.6888141) +=(0, 0.081398654) -= (0,0.081398654)
        (100, 0.792010201)+=(0, 0.073466565) -= (0, 0.073466565)
    };
    \addplot[line width=0.30mm,
    color=red,mark=x,error bars/.cd, y dir=both, y explicit
    ]
        plot coordinates {		
        	(20, 0.8884) += (0, 0.0656) -= (0,  0.0656)
     		(40, 0.7293) += (0, 0.0597) -= (0, 0.0597)
     	 (60, 0.705226009) += (0,0.065490618) -= (0,0.065490618)	
     	 (80, 0.721910058)+= (0, 0.079785695) -= (0, 0.079785695)
     	 (100, 0.760264271)+= (0,0.082013012) -= (0, 0.082013012)
        };
        \addplot[line width=0.30mm,
        color=green,mark=square,error bars/.cd, y dir=both, y explicit
        ]
        plot coordinates {			
       (20, 0.8193)+= (0,0.0808) -= (0,0.0808)
     			(40, 0.7267)+= (0, 0.0569) -= (0,0.0569)
     		 (60, 0.71659009)+= (0,0.045641364) -= (0,0.045641364)	
     		 (80,	0.689402953)+= (0,0.077690666) -= (0,0.077690666)	
     		 (100, 0.737166207)+= (0,0.094847661) -= (0,0.094847661)	
        };
      \addplot[line width=0.30mm,
        mark=*,
    cyan,
    error bars/.cd, y dir=both, y explicit,
    ] plot coordinates {
    	(20, 0.968910) += (0,0.020193) -=(0,0.020193)
(40, 0.680844) += (0,0.291579) -=(0,0.291579)
(60, 0.979862) += (0,0.017347) -=(0,0.017347)
(80, 0.967063) += (0,0.026358) -=(0,0.026358)
(100, 0.970211) += (0,0.023542) -=(0,0.023542)
    };
    \addplot[line width=0.30mm,
    color=black,mark=triangle,error bars/.cd, y dir=both, y explicit
    ]
    plot coordinates {		
    	(20, 0.751884) += (0,0.246730) -=(0,0.246730)
(40, 0.755345) += (0,0.247672) -=(0,0.247672)
(60, 0.693012) += (0,0.296514) -=(0,0.296514)
(80, 0.968239) += (0,0.025793) -=(0,0.025793)
(100, 0.990932) += (0,0.010939) -=(0,0.010939)
    };
    \addplot[line width=0.30mm,
    color=purple,mark=otimes,error bars/.cd, y dir=both, y explicit
    ]
    plot coordinates {			
    	(20, 0.861792) += (0,0.189131) -=(0,0.189131)
(40, 0.488502) += (0,0.319367) -=(0,0.319367)
(60, 0.988830) += (0,0.009351) -=(0,0.009351)
(80, 0.967587) += (0,0.025921) -=(0,0.025921)
(100, 0.991971) += (0,0.009997) -=(0,0.009997)	
    };
  \nextgroupplot[ylabel={Average Latency (ms)}, ymin=2, ymax=6, ytick={2,3,4,5},scaled y ticks = false, log ticks with fixed point={1000 sep=},
	ylabel near ticks ]
    \addplot[line width=0.30mm,
        mark=*,
        blue,
        error bars/.cd, y dir=both, y explicit,
    ] plot coordinates {	
    	(20, 3.538821) += (0,0.411912) -=(0,0.411912)
(40, 3.804337) += (0,0.344301) -=(0,0.344301)
(60, 4.388965) += (0,0.327278) -=(0,0.327278)
(80, 4.639421) += (0,0.303794) -=(0,0.303794)
(100, 4.984263) += (0,0.493649) -=(0,0.493649)
    };
    \addplot[line width=0.30mm,
    color=red,mark=x,error bars/.cd, y dir=both, y explicit
    ]
        plot coordinates {	
        	(20, 3.611643) += (0,0.391602) -=(0,0.391602)
(40, 3.832202) += (0,0.372653) -=(0,0.372653)
(60, 4.398760) += (0,0.333847) -=(0,0.333847)
(80, 4.620473) += (0,0.482547) -=(0,0.482547)
(100, 5.163448) += (0,0.423207) -=(0,0.423207)
        };
        \addplot[line width=0.30mm,
        color=green,mark=square,error bars/.cd, y dir=both, y explicit
        ]
        plot coordinates {				
      (20, 3.591603) += (0,0.412880) -=(0,0.412880)
(40, 3.965617) += (0,0.370143) -=(0,0.370143)
(60, 4.634208) += (0,0.420061) -=(0,0.420061)
(80, 4.690560) += (0,0.441071) -=(0,0.441071)
(100, 5.209910) += (0,0.499668) -=(0,0.499668)	
        };  
     
    \addplot[line width=0.30mm,
    mark=*,
    cyan,
    error bars/.cd, y dir=both, y explicit,
    ] plot coordinates {
    	(20, 2.638258) += (0,0.039341) -=(0,0.039341)
(40, 2.669051) += (0,0.029187) -=(0,0.029187)
(60, 2.756503) += (0,0.063867) -=(0,0.063867)
(80, 2.684343) += (0,0.050265) -=(0,0.050265)
(100, 2.928721) += (0,0.109843) -=(0,0.109843)
    };
    \addplot[line width=0.30mm,
    color=black,mark=triangle,error bars/.cd, y dir=both, y explicit
    ]
    plot coordinates {		
    	(20, 2.745787) += (0,0.118753) -=(0,0.118753)
(40, 2.637260) += (0,0.035007) -=(0,0.035007)
(60, 2.771339) += (0,0.109612) -=(0,0.109612)
(80, 2.695674) += (0,0.041391) -=(0,0.041391)
(100, 2.997845) += (0,0.069474) -=(0,0.069474)
    };
    \addplot[line width=0.30mm,
    color=purple,mark=otimes,error bars/.cd, y dir=both, y explicit
    ]
    plot coordinates {			
    	(20, 2.694228) += (0,0.079546) -=(0,0.079546)
(40, 2.649227) += (0,0.019153) -=(0,0.019153)
(60, 2.782981) += (0,0.101601) -=(0,0.101601)
(80, 2.696241) += (0,0.050720) -=(0,0.050720)
(100, 2.999751) += (0,0.071500) -=(0,0.071500)
    };
      \nextgroupplot[ylabel={Average Jitter (ms)}, ymin=0, ymax=0.6]
     \addplot[line width=0.30mm,
        mark=*,
        blue,
        error bars/.cd, y dir=both, y explicit,
    ] plot coordinates {
    	(20, 0.032964) += (0,0.009338) -=(0,0.009338)
(40, 0.027410) += (0,0.006685) -=(0,0.006685)
(60, 0.034219) += (0,0.009441) -=(0,0.009441)
(80, 0.030813) += (0,0.007627) -=(0,0.007627)
(100, 0.045151) += (0,0.011495) -=(0,0.011495)
    };
    \addplot[line width=0.30mm,
    color=red,mark=x,error bars/.cd, y dir=both, y explicit
    ]
        plot coordinates {		
        	(20, 0.031106) += (0,0.005890) -=(0,0.005890)
(40, 0.031090) += (0,0.009712) -=(0,0.009712)
(60, 0.039553) += (0,0.010887) -=(0,0.010887)
(80, 0.037969) += (0,0.014539) -=(0,0.014539)
(100, 0.047848) += (0,0.012639) -=(0,0.012639)
        };
        \addplot[line width=0.30mm,
        color=green,mark=square,error bars/.cd, y dir=both, y explicit
        ]
        plot coordinates {			
       (20, 0.036958) += (0,0.012901) -=(0,0.012901)
(40, 0.038244) += (0,0.011317) -=(0,0.011317)
(60, 0.052193) += (0,0.011834) -=(0,0.011834)
(80, 0.041094) += (0,0.013325) -=(0,0.013325)
(100, 0.062775) += (0,0.014470) -=(0,0.014470)
        };
      \addplot[line width=0.30mm,
    mark=*,
    cyan,
    error bars/.cd, y dir=both, y explicit,
    ] plot coordinates {
    	(20, 0.478349) += (0,0.034412) -=(0,0.034412)
(40, 0.374642) += (0,0.163151) -=(0,0.163151)
(60, 0.469852) += (0,0.026454) -=(0,0.026454)
(80, 0.467227) += (0,0.015948) -=(0,0.015948)
(100, 0.468475) += (0,0.023270) -=(0,0.023270)
    };
    \addplot[line width=0.30mm,
    color=black,mark=triangle,error bars/.cd, y dir=both, y explicit
    ]
    plot coordinates {		
    	(20, 0.407401) += (0,0.136970) -=(0,0.136970)
(40, 0.397443) += (0,0.134715) -=(0,0.134715)
(60, 0.316321) += (0,0.138187) -=(0,0.138187)
(80, 0.473500) += (0,0.020428) -=(0,0.020428)
(100, 0.462136) += (0,0.014289) -=(0,0.014289)
    };
    \addplot[line width=0.30mm,
    color=purple,mark=otimes,error bars/.cd, y dir=both, y explicit
    ]
    plot coordinates {			
    	(20, 0.459649) += (0,0.107855) -=(0,0.107855)
(40, 0.258492) += (0,0.170647) -=(0,0.170647)
(60, 0.465087) += (0,0.029030) -=(0,0.029030)
(80, 0.469949) += (0,0.012259) -=(0,0.012259)
(100, 0.459301) += (0,0.012772) -=(0,0.012772)
    };
       \end{groupplot}
       legend style={
		at={(1,1.05)},
		anchor=south east,
		column sep=1ex}
     \node (fig6_Legend ) at ($(fig6_plots c2r1.center)-(0,3cm)$){\ref{Fig6Legend}};
       \end{tikzpicture}       
       \caption{Results in Manhattan map for the 2nd proposed path selection approach }
    \label{scenario2isr}
    \end{figure*}

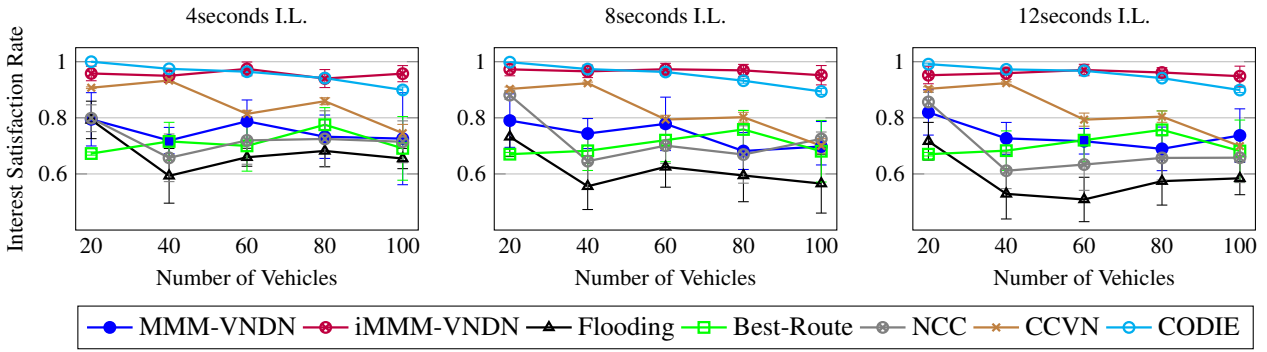
\begin{figure*}[!ht]
        	\centering
	\begin{tikzpicture}
    \begin{groupplot}[legend columns=7,legend to name= Fig3Legend,
            legend entries={{\normalsize MMM-VNDN},{\normalsize iMMM-VNDN},{\normalsize Flooding},{\normalsize Best-Route},{\normalsize NCC},{\normalsize CCVN},{\normalsize CODIE}},
            group style={group size= 3 by 1, group name = fig3_plots},
	width  =0.35\textwidth,
	height = 4cm,
	major x tick style = transparent,
	ymajorgrids = true,
	xlabel = {Number of Vehicles},
	symbolic x coords={20,40,60,80,100},
	xtick = data,
	scaled y ticks = false,
	enlarge x limits=0.05,
    ytick={0.6,0.8,1},
	ymin=0.4,
	ymax=1.05,
	ylabel near ticks]
        \nextgroupplot[title=4seconds I.L.,ylabel={Interest Satisfaction Rate }]
   \addplot[line width=0.30mm,
        mark=*,
        blue,
        error bars/.cd, y dir=both, y explicit,
    ] plot coordinates {	
    	(20,  0.7947) +=(0,  0.0950) -= (0,  0.0950)
    	(40, 0.7192) +=(0,  0.0466) -= (0, 0.0466)	
        (60, 0.787744436) +=(0, 0.075894702) -= (0, 0.075894702)		
        (80, 0.732622047) +=(0, 0.077340163) -= (0, 0.077340163)
        (100, 0.725809761)+=(0, 0.164197525) -= (0, 0.164197525)
    };
  \addplot[line width=0.30mm,color=purple,mark=otimes, error bars/.cd, y dir=both, y explicit]
  plot coordinates {	
  	(20, 0.958060) += (0,0.025884) -=(0,0.025884)
(40, 0.949685) += (0,0.023541) -=(0,0.023541)
(60, 0.974407) += (0,0.024529) -=(0,0.024529)
(80, 0.939927) += (0,0.032093) -=(0,0.032093)
(100, 0.957353) += (0,0.028792) -=(0,0.028792)
  };
	\addplot[line width=0.30mm,
              color=black,mark=triangle,
        error bars/.cd, y dir=both, y explicit,
    ] plot coordinates {			
    	(20,0.79258)+= (0,0.066762) -= (0,0.066762)
    	(40, 0.593285 )+= (0,0.097474) -= (0,0.097474)
	    (60, 0.659465) +=(0, 0.032010) -= (0, 0.032010)		
        (80,  0.681982) +=(0, 0.055543) -= (0,0.055543)
        (100, 0.654944)+=(0, 0.036107) -= (0, 0.036107)
    };
	\addplot[line width=0.30mm,
	color=green,mark=square,
	error bars/.cd, y dir=both, y explicit,
	] plot coordinates {			
		(20, 0.6727)+= (0, 0.0057) -= (0, 0.0057)
		(40, 0.7159 )+= (0, 0.0680) -= (0, 0.0680)
		(60, 0.7000) +=(0, 0.0901) -= (0, 0.0901)		
		(80,   0.7753) +=(0, 0.0610) -= (0,0.0610)
		(100, 0.6911) +=(0, 0.1135) -= (0, 0.1135)
	};
	\addplot[line width=0.30mm,
	color=gray,mark=otimes,
	error bars/.cd, y dir=both, y explicit,
	] plot coordinates {			
		(20,0.798605)+= (0, 0.047721) -= (0, 0.047721)
		(40, 0.657504 )+= (0,0.084857) -= (0,0.084857)
		(60, 0.7195) +=(0,  0.0849) -= (0,  0.0849)		
		(80,  0.7247) +=(0, 0.100320) -= (0,0.100320)
		(100, 0.7146)+=(0, 0.074542) -= (0, 0.074542)
	};
    \addplot[line width=0.30mm,
	color=brown,mark=x,
	error bars/.cd, y dir=both, y explicit,
	] plot coordinates {			
		(20, 0.907251) += (0,0.003064) -=(0,0.003064)
(40, 0.932910) += (0,0.003322) -=(0,0.003322)
(60, 0.814031) += (0,0.012703) -=(0,0.012703)
(80, 0.859169) += (0,0.012366) -=(0,0.012366)
(100, 0.745291) += (0,0.031247) -=(0,0.031247)
	};
\addplot[line width=0.30mm,
	color=cyan,mark=o,
	error bars/.cd, y dir=both, y explicit,
	] plot coordinates {			
		(20, 1.000000) += (0,0.000000) -=(0,0.000000)
(40, 0.974619) += (0,0.001211) -=(0,0.001211)
(60, 0.964342) += (0,0.002213) -=(0,0.002213)
(80, 0.941607) += (0,0.005726) -=(0,0.005726)
(100, 0.899354) += (0,0.010578) -=(0,0.010578)
	};
	   	]  \nextgroupplot[title=8seconds I.L.]
    \addplot[mark=*,	line width=0.30mm,blue,  error bars/.cd, y dir=both, y explicit]
     	plot coordinates {	
     		(20, 0.7902) +=(0, 0.0954) -= (0, 0.0954)
        	(40, 0.7440) +=(0,  0.0534) -= (0, 0.0534)	
            (60, 0.77848) += (0,0.095223428) -= (0,0.095223428)	
            (80, 0.681671)+= (0, 0.065734716) -= (0, 0.065734716)
            (100, 0.697776)+= (0,0.08909202) -= (0, 0.065734716)
     	};
   	 \addplot[line width=0.30mm,color=purple,mark=otimes,error bars/.cd, y dir=both, y explicit]
     	 plot coordinates {	
     	 	(20, 0.972748) += (0,0.022961) -=(0,0.022961)
(40, 0.965608) += (0,0.022282) -=(0,0.022282)
(60, 0.973100) += (0,0.021115) -=(0,0.021115)
(80, 0.969130) += (0,0.021302) -=(0,0.021302)
(100, 0.951852) += (0,0.034469) -=(0,0.034469)
     	 };
   	    \addplot[line width=0.30mm,color=black,mark=triangle,error bars/.cd, y dir=both, y explicit]
     	    plot coordinates {		
     	    		(20, 0.7323)+= (0,0.0704) -= (0, 0.0704)
     	    		(40, 0.5560 )+= (0,0.0828) -= (0,0.0828)
     	    		(60, 0.6251) +=(0, 0.0723) -= (0, 0.0723)		
     	        	(80, 0.5945) +=(0,  0.0935) -= (0, 0.0935)
     	    	(100, 0.5656) += (0,  0.1051) -= (0,   0.1051)
     	    };
     	\addplot[
     	line width=0.30mm,
     	color=green,mark=square,
     	error bars/.cd, y dir=both, y explicit,
     	] plot coordinates {			
     		(20,  0.6704)+= (0, 0.0063) -= (0,0.0063)
     		(40,  0.6825 )+= (0, 0.0702) -= (0, 0.0702)
     		(60, 0.7204) +=(0,  0.0769) -= (0, 0.0769)		
     		(80,  0.7585) +=(0, 0.0678) -= (0, 0.0678)
     		(100,  0.6809) +=(0, 0.1097) -= (0,0.1097)
     	};
     	\addplot[
     	line width=0.30mm,
     	color=gray,mark=otimes,
     	error bars/.cd, y dir=both, y explicit,
     	] plot coordinates {			
     		(20,0.880855)+= (0,0.015599) -= (0,0.015599)
     		(40, 0.645622 )+= (0,0.098803) -= (0,0.098803)
     		(60, 0.700820) +=(0, 0.063414) -= (0, 0.063414)		
     		(80,  0.669246) +=(0, 0.102167) -= (0,0.102167)
     		(100, 0.7249)+=(0, 0.024520) -= (0, 0.024520)
     	};
     	  \addplot[line width=0.30mm,
		color=brown,mark=x,
	error bars/.cd, y dir=both, y explicit,
	] plot coordinates {			
(20, 0.902944) += (0,0.007016) -=(0,0.007016)
(40, 0.923284) += (0,0.004137) -=(0,0.004137)
(60, 0.794046) += (0,0.015003) -=(0,0.015003)
(80, 0.802215) += (0,0.017542) -=(0,0.017542)
(100, 0.700324) += (0,0.025342) -=(0,0.025342)
	};
\addplot[line width=0.30mm,
	color=cyan,mark=o,
	error bars/.cd, y dir=both, y explicit,
	] plot coordinates {			
		(20, 0.998300) += (0,0.000829) -=(0,0.000829)
(40, 0.973693) += (0,0.002281) -=(0,0.002281)
(60, 0.963394) += (0,0.004163) -=(0,0.004163)
(80, 0.932274) += (0,0.008848) -=(0,0.008848)
(100, 0.893960) += (0,0.010563) -=(0,0.010563)
	};
	             \nextgroupplot[title=12seconds I.L.]
  		\addplot[line width=0.30mm,
  		color=blue,mark=*,error bars/.cd, y dir=both, y explicit
  		]
     		plot coordinates {	
     			(20, 0.8193)+= (0,0.0808) -= (0,0.0808)
     			(40, 0.7267)+= (0, 0.0569) -= (0,0.0569)
     		 (60, 0.71659009)+= (0,0.045641364) -= (0,0.045641364)	
     		 (80,	0.689402953)+= (0,0.077690666) -= (0,0.077690666)	
     		 (100, 0.737166207)+= (0,0.094847661) -= (0,0.094847661)	
     		};
  	   \addplot[line width=0.30mm,
  	   color=purple,mark=otimes,error bars/.cd, y dir=both, y explicit
  	   ]
     		   plot coordinates {	
     		(20, 0.951730) += (0,0.030339) -=(0,0.030339)
(40, 0.959032) += (0,0.021786) -=(0,0.021786)
(60, 0.970428) += (0,0.020318) -=(0,0.020318)
(80, 0.961571) += (0,0.017319) -=(0,0.017319)
(100, 0.948385) += (0,0.035764) -=(0,0.035764)
     		   };
          \addplot[line width=0.30mm,
          color=black,mark=triangle,error bars/.cd, y dir=both, y explicit
          ]
     		      plot coordinates {	
     	    	(20, 0.7165)+= (0,   0.0672) -= (0,   0.0672)
     	       	(40, 0.5290)+= (0,0.0895) -= (0,0.0895)
     	      	(60, 0.5090) += (0,0.0791) -= (0,0.0791)	
     	      	(80, 0.5743) += (0, 0.0853) -= (0, 0.0853)	
     		   	(100, 0.5845)+= (0, 0.0585) -= (0, 0.0585)	
     				        };
           	\addplot[
        	line width=0.30mm,color=green, mark=square,   				        	error bars/.cd, y dir=both, y explicit,] 
            plot coordinates {			
          		(20,  0.6703)+= (0, 0.0063) -= (0,0.0063)
         		(40, 0.6830 )+= (0,  0.0704) -= (0,  0.0704)
           		(60, 0.7204) +=(0,0.0769) -= (0,0.0769)		
          		(80,   0.7571) +=(0, 0.0675) -= (0, 0.0675)
        		(100,   0.6818) +=(0,  0.1103) -= (0,  0.1103)
     				        	};
      	\addplot[line width=0.30mm,
     				        	color=gray,mark=otimes,
     				        	error bars/.cd, y dir=both, y explicit,
 	        	] plot coordinates {			
 	        		(20,0.856617)+= (0,0.033899) -= (0,0.033899)
 	        		(40, 0.611421 )+= (0,0.063798) -= (0,0.063798)
           		(60, 0.633484) +=(0,0.091911) -= (0, 0.091911)		
           		(80,  0.657113) +=(0, 0.078879) -= (0,0.078879)
           		(100, 0.657864)+=(0, 0.088653) -= (0, 0.088653)
     				        	};
     		  \addplot[line width=0.30mm,
		color=brown,mark=x,
	error bars/.cd, y dir=both, y explicit,
	] plot coordinates {			
	(20, 0.903340) += (0,0.007290) -=(0,0.007290)
(40, 0.923616) += (0,0.004383) -=(0,0.004383)
(60, 0.793517) += (0,0.023393) -=(0,0.023393)
(80, 0.804299) += (0,0.020241) -=(0,0.020241)
(100, 0.699280) += (0,0.037631) -=(0,0.037631)
	};
\addplot[line width=0.30mm,
	color=cyan,mark=o,
	error bars/.cd, y dir=both, y explicit,
	] plot coordinates {			
		(20, 0.991266) += (0,0.003556) -=(0,0.003556)
(40, 0.972672) += (0,0.001984) -=(0,0.001984)
(60, 0.967640) += (0,0.003232) -=(0,0.003232)
(80, 0.941518) += (0,0.007468) -=(0,0.007468)
(100, 0.898947) += (0,0.008979) -=(0,0.008979)
	};
       \end{groupplot}
       legend style={
		at={(1,1.05)},
		anchor=south east,
		column sep=1ex}
     \node (fig3_Legend) at ($(fig3_plots c2r1.center)-(0,2.5cm)$){\ref{Fig3Legend}};
       \end{tikzpicture}       
       \caption{Interest Satisfaction Rate in Manhattan map for the 3rd  proposed path selection approach}
       \label{scenario3isr}
    \end{figure*} 
    
          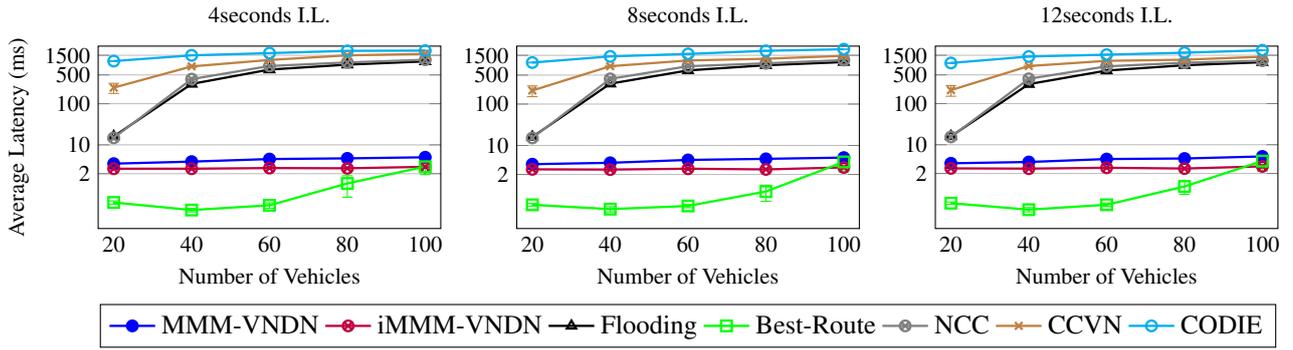
\begin{figure*}[!ht]
        	\centering
	\begin{tikzpicture}
 \begin{groupplot}[legend columns=7,legend to name= Fig4Legend,
            legend entries={{\normalsize MMM-VNDN},{\normalsize iMMM-VNDN},{\normalsize Flooding},{\normalsize Best-Route},{\normalsize NCC},{\normalsize CCVN},{\normalsize CODIE}},
            group style={group size= 3 by 1, group name = fig4_plots},
	width  =0.35\textwidth,
	height = 4cm,
	major x tick style = transparent,
	ymajorgrids = true,
	xlabel = {Number of Vehicles},
	symbolic x coords={20,40,60,80,100},
	xtick = data,
	scaled y ticks = false,
    ytick={0,2,10,100,500,1500},
	enlarge x limits=0.05,
	ymin=0,
	ymax=2500,
    log ticks with fixed point={1000 sep=},
	ylabel near ticks	]
        \nextgroupplot[ymode=log,title=4seconds I.L.,ylabel={Average Latency (ms) }]
     		]
            \addplot[line width=0.30mm,
  		color=blue,mark=*,error bars/.cd, y dir=both, y explicit
  		]
  	plot coordinates {
  		(20, 3.501575) += (0,0.427472) -=(0,0.427472)
  		(40, 3.921669) += (0,0.331418) -=(0,0.331418)
  		(60, 4.525261) += (0,0.353064) -=(0,0.353064)
  		(80, 4.702927) += (0,0.302383) -=(0,0.302383)
  		(100, 4.996934) += (0,0.372611) -=(0,0.372611)
  	};
  	\addplot[line width=0.30mm,
  	color=purple,mark=otimes,error bars/.cd, y dir=both, y explicit
  	]
  	plot coordinates {	
  	(20, 2.633279) += (0,0.035620) -=(0,0.035620)
  	(40, 2.635812) += (0,0.032892) -=(0,0.032892)
  	(60, 2.734655) += (0,0.065073) -=(0,0.065073)
  	(80, 2.699700) += (0,0.054017) -=(0,0.054017)
  	(100, 2.930934) += (0,0.120794) -=(0,0.120794)
  	};
  	\addplot[line width=0.30mm,
  	color=black,mark=triangle,
  	error bars/.cd, y dir=both, y explicit]
  	plot coordinates {	
  		(20, 16.048110) += (0,0.249602) -=(0,0.249602)
  		(40, 302.001800) += (0,8.997991) -=(0,8.997991)
  		(60, 673.745500) += (0,14.691194) -=(0,14.691194)
  		(80, 887.997600) += (0,14.639898) -=(0,14.639898)
  		(100, 1065.963000) += (0,24.909943) -=(0,24.909943)
  	};
  	\addplot[line width=0.30mm,
  	color=green,mark=square,
  	error bars/.cd, y dir=both, y explicit] 
  	plot coordinates {			
  		(20, 0.399327) += (0,0.036979) -=(0,0.036979)
  		(40, 0.261297) += (0,0.026226) -=(0,0.026226)
  		(60, 0.340597) += (0,0.050608) -=(0,0.050608)
  		(80, 1.158518) += (0,0.622201) -=(0,0.622201)
  		(100, 2.952761) += (0,1.048438) -=(0,1.048438)
  	};
  	\addplot[line width=0.30mm,
  	color=gray,mark=otimes,
  	error bars/.cd, y dir=both, y explicit] 
  	plot coordinates {			
  		(20, 14.813860) += (0,0.274972) -=(0,0.274972)
  		(40, 395.808400) += (0,13.895766) -=(0,13.895766)
  		(60, 823.119000) += (0,12.960140) -=(0,12.960140)
  		(80, 1003.609000) += (0,22.194771) -=(0,22.194771)
  		(100, 1166.357000) += (0,16.113241) -=(0,16.113241)
  	};
      \addplot[line width=0.30mm,
	color=brown,mark=x,
	error bars/.cd, y dir=both, y explicit,
	] plot coordinates {			
	(20, 245.435200) += (0,67.596293) -=(0,67.596293)
	(40, 804.002300) += (0,45.060141) -=(0,45.060141)
	(60, 1154.189000) += (0,34.582704) -=(0,34.582704)
	(80, 1492.194000) += (0,75.440319) -=(0,75.440319)
	(100, 1620.556000) += (0,85.731005) -=(0,85.731005)
	};
\addplot[line width=0.30mm,
	color=cyan,mark=o,
	error bars/.cd, y dir=both, y explicit,
	] plot coordinates {			
	(20, 1083.290000) += (0,18.853985) -=(0,18.853985)
(40, 1491.662000) += (0,10.936775) -=(0,10.936775)
(60, 1692.865000) += (0,19.509217) -=(0,19.509217)
(80, 1920.538000) += (0,23.433382) -=(0,23.433382)
(100, 1947.033000) += (0,32.485453) -=(0,32.485453)
	};
     \nextgroupplot[ymode=log,title=8seconds I.L.]
         	\addplot[line width=0.30mm,
         color=blue,mark=*,error bars/.cd, y dir=both, y explicit
         	]
     	plot coordinates {
     		(20, 3.533157) += (0,0.377173) -=(0,0.377173)
     		(40, 3.820077) += (0,0.381719) -=(0,0.381719)
     		(60, 4.477183) += (0,0.286628) -=(0,0.286628)
     		(80, 4.754829) += (0,0.373236) -=(0,0.373236)
     		(100, 5.097722) += (0,0.331335) -=(0,0.331335)
     	};
    \addplot[line width=0.30mm,
    color=purple,mark=otimes,error bars/.cd, y dir=both, y explicit
    ]
    plot coordinates {	
    (20, 2.656397) += (0,0.038817) -=(0,0.038817)
    (40, 2.620867) += (0,0.029123) -=(0,0.029123)
    (60, 2.747503) += (0,0.067443) -=(0,0.067443)
    (80, 2.645608) += (0,0.039820) -=(0,0.039820)
    (100, 2.952085) += (0,0.078282) -=(0,0.078282)
    };
    \addplot[line width=0.30mm,
    color=black,mark=triangle,error bars/.cd, y dir=both, y explicit
    ]
    plot coordinates {	
    	(20, 16.049360) += (0,0.230488) -=(0,0.230488)
    	(40, 310.315200) += (0,8.030360) -=(0,8.030360)
    	(60, 648.695500) += (0,12.839197) -=(0,12.839197)
    	(80, 865.545800) += (0,12.619952) -=(0,12.619952)
    	(100, 1030.876000) += (0,33.055213) -=(0,33.055213)
    };
     \addplot[line width=0.30mm,
    color=green,mark=square,
    error bars/.cd, y dir=both, y explicit,
    ] plot coordinates {			
    	(20, 0.369993) += (0,0.026765) -=(0,0.026765)
    	(40, 0.290748) += (0,0.042083) -=(0,0.042083)
    	(60, 0.346671) += (0,0.033003) -=(0,0.033003)
    	(80, 0.774609) += (0,0.328164) -=(0,0.328164)
    	(100, 4.004968) += (0,1.707050) -=(0,1.707050)
    };
    \addplot[line width=0.30mm,
    color=gray,mark=otimes,
    error bars/.cd, y dir=both, y explicit,
    ] plot coordinates {			
    	(20, 15.206950) += (0,0.249067) -=(0,0.249067)
    	(40, 405.846800) += (0,12.057755) -=(0,12.057755)
    	(60, 818.704700) += (0,12.325203) -=(0,12.325203)
    	(80, 960.853300) += (0,29.394156) -=(0,29.394156)
    	(100, 1151.166000) += (0,11.211593) -=(0,11.211593)
    };
      \addplot[line width=0.30mm,
	color=brown,mark=x,
	error bars/.cd, y dir=both, y explicit,
	] plot coordinates {			
	(20, 212.228900) += (0,61.563099) -=(0,61.563099)
	(40, 818.764300) += (0,33.936107) -=(0,33.936107)
	(60, 1126.501000) += (0,47.124645) -=(0,47.124645)
	(80, 1240.222000) += (0,36.834393) -=(0,36.834393)
	(100, 1457.462000) += (0,86.717906) -=(0,86.717906)
	};
\addplot[line width=0.30mm,
	color=cyan,mark=o,
	error bars/.cd, y dir=both, y explicit,
	] plot coordinates {			
	(20, 1003.719600) += (0,24.419981) -=(0,24.419981)
(40, 1411.856000) += (0,19.334218) -=(0,19.334218)
(60, 1621.108000) += (0,28.478864) -=(0,28.478864)
(80, 1927.110000) += (0,30.926531) -=(0,30.926531)
(100, 2105.514000) += (0,22.726584) -=(0,22.726584)
	};
    \nextgroupplot[ymode=log,title=12seconds I.L.]	
       	\addplot[line width=0.30mm,
       	color=blue,mark=*,error bars/.cd, y dir=both, y explicit
       	]
      	plot coordinates {	
      	(20, 3.536597) += (0,0.383972) -=(0,0.383972)
      	(40, 3.803727) += (0,0.358291) -=(0,0.358291)
      	(60, 4.486650) += (0,0.460240) -=(0,0.460240)
      	(80, 4.625242) += (0,0.385947) -=(0,0.385947)
      	(100, 5.186654) += (0,0.470767) -=(0,0.470767)
      	};  
      	 \addplot[line width=0.30mm,
      	 color=purple,mark=otimes,error bars/.cd, y dir=both, y explicit
      	 ]
      	 plot coordinates {	
      	(20, 2.659859) += (0,0.035318) -=(0,0.035318)
      	(40, 2.619455) += (0,0.029371) -=(0,0.029371)
      	(60, 2.753552) += (0,0.069825) -=(0,0.069825)
      	(80, 2.636086) += (0,0.041552) -=(0,0.041552)
      	(100, 2.971773) += (0,0.074466) -=(0,0.074466)
      	 };
      	 \addplot[line width=0.30mm,
      	 color=black,mark=triangle,error bars/.cd, y dir=both, y explicit
      	 ]
      	 plot coordinates {	
      	 (20, 15.994990) += (0,0.395964) -=(0,0.395964)
      	 (40, 297.636100) += (0,12.489406) -=(0,12.489406)
      	 (60, 636.840600) += (0,19.915761) -=(0,19.915761)
      	 (80, 858.898900) += (0,14.856418) -=(0,14.856418)
      	 (100, 1016.485300) += (0,14.143498) -=(0,14.143498)
      	 };
      	 \addplot[line width=0.30mm,
      	 color=green,mark=square,
      	 error bars/.cd, y dir=both, y explicit,
      	 ] plot coordinates {			
      	 (20, 0.379175) += (0,0.026395) -=(0,0.026395)
      	 (40, 0.263345) += (0,0.031471) -=(0,0.031471)
      	 (60, 0.346671) += (0,0.033003) -=(0,0.033003)
      	 (80, 0.964743) += (0,0.347112) -=(0,0.347112)
      	 (100, 4.004968) += (0,1.707050) -=(0,1.707050)
      	 };
      	 \addplot[line width=0.30mm,
      	 color=gray,mark=otimes,
      	 error bars/.cd, y dir=both, y explicit,
      	 ] plot coordinates {			
      	 (20, 15.469300) += (0,0.510004) -=(0,0.510004)
      	 (40, 401.703300) += (0,15.238161) -=(0,15.238161)
      	 (60, 805.519300) += (0,10.213196) -=(0,10.213196)
      	 (80, 985.654700) += (0,13.141832) -=(0,13.141832)
      	 (100, 1104.649000) += (0,22.054914) -=(0,22.054914)
      	 };
       \addplot[line width=0.30mm,
	color=brown,mark=x,
	error bars/.cd, y dir=both, y explicit,
	] plot coordinates {			
	(20, 212.589800) += (0,61.442992) -=(0,61.442992)
	(40, 819.368500) += (0,34.276953) -=(0,34.276953)
	(60, 1097.156900) += (0,44.113858) -=(0,44.113858)
	(80, 1168.304700) += (0,48.918036) -=(0,48.918036)
	(100, 1379.736000) += (0,70.516081) -=(0,70.516081)
	};
\addplot[line width=0.30mm,
	color=cyan,mark=o,
	error bars/.cd, y dir=both, y explicit,
	] plot coordinates {			
	(20, 971.781800) += (0,24.080546) -=(0,24.080546)
(40, 1396.183000) += (0,26.888420) -=(0,26.888420)
(60, 1548.890000) += (0,18.637411) -=(0,18.637411)
(80, 1731.143000) += (0,25.914659) -=(0,25.914659)
(100, 1980.440000) += (0,25.175269) -=(0,25.175269)
	};
     \end{groupplot}
       legend style={
		at={(1,1.05)},
		anchor=south east,
		column sep=1ex}
     \node (fig4_Legend ) at ($(fig4_plots c2r1.center)-(0,2.5cm)$){\ref{Fig4Legend}};
       \end{tikzpicture}       
       \caption{Average Latency in Manhattan map for the 3rd proposed path selection approach}
       \label{scenario3ltn}
    \end{figure*}   
    
     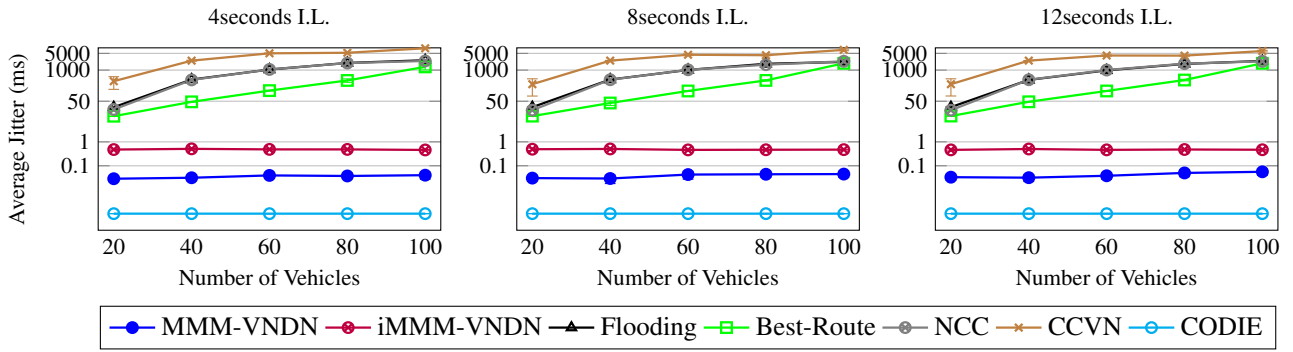
\begin{figure*}[!ht]
        	\centering
	\begin{tikzpicture}
 \begin{groupplot}[legend columns=7,legend to name= Fig5Legend,
            legend entries={{\normalsize MMM-VNDN},{\normalsize iMMM-VNDN},{\normalsize Flooding},{\normalsize Best-Route},{\normalsize NCC},{\normalsize CCVN},{\normalsize CODIE}},
            group style={group size= 3 by 1, group name = fig5_plots},
	width  =0.35\textwidth,
	height = 4cm,
	major x tick style = transparent,
	ymajorgrids = true,
	xlabel = {Number of Vehicles},
	symbolic x coords={20,40,60,80,100},
	xtick = data,
	scaled y ticks = false,
	enlarge x limits=0.05,
	ymin=-1,
	ymax=8500,
    ymode=log,
     ytick={0,0.1,1,50,1000,5000},
    log ticks with fixed point={1000 sep=},
	ylabel near ticks	]
        \nextgroupplot[title=4seconds I.L.,ylabel={Average Jitter (ms) }]
     		]
            \addplot[line width=0.30mm,
  		color=blue,mark=*,error bars/.cd, y dir=both, y explicit
  		]
  	plot coordinates {
  		(20, 0.028830) += (0,0.005889) -=(0,0.005889)
(40, 0.031654) += (0,0.007458) -=(0,0.007458)
(60, 0.039606) += (0,0.006477) -=(0,0.006477)
(80, 0.037584) += (0,0.007270) -=(0,0.007270)
(100, 0.040523) += (0,0.013291) -=(0,0.013291)
  	};
  	\addplot[line width=0.30mm,
  	color=purple,mark=otimes,error bars/.cd, y dir=both, y explicit
  	]
  	plot coordinates {	
  		(20, 0.477091) += (0,0.030007) -=(0,0.030007)
(40, 0.510612) += (0,0.047666) -=(0,0.047666)
(60, 0.485125) += (0,0.027203) -=(0,0.027203)
(80, 0.483857) += (0,0.013567) -=(0,0.013567)
(100, 0.459748) += (0,0.015944) -=(0,0.015944)
  	};
  	\addplot[line width=0.30mm,
  	color=black,mark=triangle,
  	error bars/.cd, y dir=both, y explicit]
  	plot coordinates {	
  		(20, 28.086050) += (0,0.546358) -=(0,0.546358)
(40, 401.701200) += (0,6.971467) -=(0,6.971467)
(60, 1053.275000) += (0,15.715518) -=(0,15.715518)
(80, 1988.236000) += (0,25.110796) -=(0,25.110796)
(100, 2575.891000) += (0,31.677306) -=(0,31.677306)
  	};
  	\addplot[line width=0.30mm,
  	color=green,mark=square,
  	error bars/.cd, y dir=both, y explicit] 
  	plot coordinates {			
  		(20, 11.800740) += (0,0.256539) -=(0,0.256539)
(40, 47.044610) += (0,0.726859) -=(0,0.726859)
(60, 138.655100) += (0,4.643980) -=(0,4.643980)
(80, 368.520500) += (0,60.332468) -=(0,60.332468)
(100, 1359.352600) += (0,344.888152) -=(0,344.888152)
  	};
  	\addplot[line width=0.30mm,
  	color=gray,mark=otimes,
  	error bars/.cd, y dir=both, y explicit] 
  	plot coordinates {			
  		(20, 22.543060) += (0,0.523227) -=(0,0.523227)
(40, 388.126900) += (0,13.847357) -=(0,13.847357)
(60, 1062.127000) += (0,8.411889) -=(0,8.411889)
(80, 1966.096000) += (0,13.550159) -=(0,13.550159)
(100, 2403.164000) += (0,33.895878) -=(0,33.895878)
  	};
      \addplot[line width=0.30mm,
	color=brown,mark=x,
	error bars/.cd, y dir=both, y explicit,
	] plot coordinates {			
		(20, 343.071800) += (0,186.420668) -=(0,186.420668)
(40, 2479.069000) += (0,115.430696) -=(0,115.430696)
(60, 5014.775000) += (0,164.990001) -=(0,164.990001)
(80, 5292.774000) += (0,102.462679) -=(0,102.462679)
(100, 8123.749000) += (0,326.208024) -=(0,326.208024)
	};
\addplot[line width=0.30mm,
	color=cyan,mark=o,
	error bars/.cd, y dir=both, y explicit,
	] plot coordinates {			
		(20,0.001)+= (0,0) -= (0,0)
		(40,0.001)+= (0,0) -= (0,0)
		(60,0.001) +=(0,0) -= (0,0)		
		(80,0.0010) +=(0,0) -= (0,0)
		(100,0.0010)+=(0,0) -= (0,0)
	};
     \nextgroupplot[title=8seconds I.L.]
         	\addplot[line width=0.30mm,
         	color=blue,mark=*,error bars/.cd, y dir=both, y explicit
         	]
     	plot coordinates {
     		(20, 0.030587) += (0,0.006662) -=(0,0.006662)
(40, 0.029445) += (0,0.011380) -=(0,0.011380)
(60, 0.043015) += (0,0.015534) -=(0,0.015534)
(80, 0.044264) += (0,0.011586) -=(0,0.011586)
(100, 0.045042) += (0,0.012795) -=(0,0.012795)
     	};
    \addplot[line width=0.30mm,
    color=purple,mark=otimes,error bars/.cd, y dir=both, y explicit
    ]
    plot coordinates {	
    	(20, 0.492584) += (0,0.056545) -=(0,0.056545)
(40, 0.505637) += (0,0.032633) -=(0,0.032633)
(60, 0.462266) += (0,0.018423) -=(0,0.018423)
(80, 0.468499) += (0,0.017091) -=(0,0.017091)
(100, 0.473533) += (0,0.018004) -=(0,0.018004)
    };
    \addplot[line width=0.30mm,
    color=black,mark=triangle,error bars/.cd, y dir=both, y explicit
    ]
    plot coordinates {	
    	(20, 28.237810) += (0,0.477390) -=(0,0.477390)
(40, 401.444900) += (0,11.460156) -=(0,11.460156)
(60, 1013.922900) += (0,20.082427) -=(0,20.082427)
(80, 1829.814000) += (0,20.949562) -=(0,20.949562)
(100, 2191.499000) += (0,477.828306) -=(0,477.828306)
    };
     \addplot[line width=0.30mm,
    color=green,mark=square,
    error bars/.cd, y dir=both, y explicit,
    ] plot coordinates {			
    	(20, 11.902120) += (0,0.349382) -=(0,0.349382)
(40, 42.080730) += (0,9.213608) -=(0,9.213608)
(60, 133.922100) += (0,4.392420) -=(0,4.392420)
(80, 369.734400) += (0,22.421519) -=(0,22.421519)
(100, 1920.237600) += (0,407.150110) -=(0,407.150110)
    };
    \addplot[line width=0.30mm,
    color=gray,mark=otimes,
    error bars/.cd, y dir=both, y explicit,
    ] plot coordinates {			
    	(20, 21.157140) += (0,0.711155) -=(0,0.711155)
(40, 394.719600) += (0,11.361171) -=(0,11.361171)
(60, 1027.043000) += (0,12.167675) -=(0,12.167675)
(80, 1678.808000) += (0,366.201384) -=(0,366.201384)
(100, 2327.392000) += (0,17.303907) -=(0,17.303907)
    };
      \addplot[line width=0.30mm,
	color=brown,mark=x,
	error bars/.cd, y dir=both, y explicit,
	] plot coordinates {			
		(20, 256.860700) += (0,174.740111) -=(0,174.740111)
(40, 2486.891000) += (0,46.623203) -=(0,46.623203)
(60, 4369.114000) += (0,325.139799) -=(0,325.139799)
(80, 4181.371000) += (0,205.270364) -=(0,205.270364)
(100, 6980.404000) += (0,509.630383) -=(0,509.630383)
	};
\addplot[line width=0.30mm,
	color=cyan,mark=o,
	error bars/.cd, y dir=both, y explicit,
	] plot coordinates {			
		(20,0.001)+= (0, 0) -= (0, 0)
		(40, 0.001)+= (0,0) -= (0,0)
		(60, 0.001) +=(0,  0) -= (0,  0)		
		(80, 0.001) +=(0, 0) -= (0,0)
		(100, 0.001)+=(0, 0) -= (0, 0)
	};
    \nextgroupplot[title=12seconds I.L.]	
       	\addplot[line width=0.30mm,
       	color=blue,mark=*,error bars/.cd, y dir=both, y explicit
       	]
      	plot coordinates {	
      		(20, 0.033415) += (0,0.007851) -=(0,0.007851)
(40, 0.031899) += (0,0.009945) -=(0,0.009945)
(60, 0.038292) += (0,0.010763) -=(0,0.010763)
(80, 0.050312) += (0,0.015340) -=(0,0.015340)
(100, 0.056169) += (0,0.016018) -=(0,0.016018)
      	};  
      	 \addplot[line width=0.30mm,
      	 color=purple,mark=otimes,error bars/.cd, y dir=both, y explicit
      	 ]
      	 plot coordinates {	
      	 (20, 0.462613) += (0,0.047382) -=(0,0.047382)
(40, 0.503169) += (0,0.032366) -=(0,0.032366)
(60, 0.463663) += (0,0.017928) -=(0,0.017928)
(80, 0.479586) += (0,0.020127) -=(0,0.020127)
(100, 0.468651) += (0,0.020267) -=(0,0.020267)
      	 };
      	 \addplot[line width=0.30mm,
      	 color=black,mark=triangle,error bars/.cd, y dir=both, y explicit
      	 ]
      	 plot coordinates {	
      	 (20, 28.338820) += (0,0.573708) -=(0,0.573708)
(40, 389.098100) += (0,14.349804) -=(0,14.349804)
(60, 974.557100) += (0,23.150192) -=(0,23.150192)
(80, 1810.363000) += (0,30.066456) -=(0,30.066456)
(100, 2410.526000) += (0,21.457158) -=(0,21.457158)
      	 };
      	 \addplot[line width=0.30mm,
      	 color=green,mark=square,
      	 error bars/.cd, y dir=both, y explicit,
      	 ] plot coordinates {			
      	(20, 12.001380) += (0,0.308220) -=(0,0.308220)
(40, 47.329260) += (0,0.781405) -=(0,0.781405)
(60, 133.922100) += (0,4.392420) -=(0,4.392420)
(80, 384.053400) += (0,27.947234) -=(0,27.947234)
(100, 1920.237600) += (0,407.150110) -=(0,407.150110)
      	 };
      	 \addplot[line width=0.30mm,
      	 color=gray,mark=otimes,
      	 error bars/.cd, y dir=both, y explicit,
      	 ] plot coordinates {			
      	 (20, 21.027700) += (0,0.441839) -=(0,0.441839)
(40, 394.475400) += (0,11.051711) -=(0,11.051711)
(60, 1005.243300) += (0,14.204328) -=(0,14.204328)
(80, 1854.742000) += (0,19.486628) -=(0,19.486628)
(100, 2321.223000) += (0,14.163926) -=(0,14.163926)
      	 };
       \addplot[line width=0.30mm,
	color=brown,mark=x,
	error bars/.cd, y dir=both, y explicit,
	] plot coordinates {			
		(20, 257.244200) += (0,174.617748) -=(0,174.617748)
(40, 2489.477000) += (0,46.241813) -=(0,46.241813)
(60, 4064.259000) += (0,245.906755) -=(0,245.906755)
(80, 4021.320000) += (0,231.045354) -=(0,231.045354)
(100, 6228.031000) += (0,478.734088) -=(0,478.734088)
	};
\addplot[line width=0.30mm,
	color=cyan,mark=o,
	error bars/.cd, y dir=both, y explicit,
	] plot coordinates {			
		(20,0.001)+= (0, 0) -= (0, 0)
		(40, 0.001)+= (0,0) -= (0,0)
		(60, 0.001) +=(0,  0) -= (0,  0)		
		(80, 0.001) +=(0, 0) -= (0,0)
		(100, 0.001)+=(0, 0) -= (0, 0)
	};
     \end{groupplot}
       legend style={
		at={(1,1.05)},
		anchor=south east,
		column sep=1ex}
     \node (fig5_Legend ) at ($(fig5_plots c2r1.center)-(0,2.5cm)$){\ref{Fig4Legend}};
       \end{tikzpicture} 
       \caption{Average Jitter in Manhattan map for the 3rd proposed path selection approach}
       \label{scenario3jt}
    \end{figure*}   
\subsection{Simulation Parameters}
We used the ndnSIM simulator \cite{afanasyev2012ndnSIM} as our evaluation environment. NdnSIM is a software module providing the basic NDN implementation \cite{afanasyev2012ndnSIM,mastorakis2016ndnSIM,afanasyev2016devGuide} for the ns-3 network simulator \cite{networkSimulator}. 

To obtain the network traffic simulation we used SUMO \cite{behrisch2011sumo}. We have chosen two different topologies to evaluate the algorithms, the Manhattan map, and the Luxembourg map \cite{codeca2015luxembourg}. The Manhattan map is a 1km x 1km grid, with the number of nodes (cars) varying from 20 to 100 and the average speed is around 15m/s. In the Luxembourg map, we have chosen an area of 1km x 1km in the city center. Luxembourg traces are available for 24 hours. We have chosen different times during these 24 hours to extract the mobility traces. Then, in these different mobility traces, we have extracted the density of vehicles, varying from 109 to 396. The average speed of cars depends on the time slot when the mobility traces were extracted. The parameters of the algorithms are shown in Table~\ref{table2}. All nodes in all algorithms have three network interfaces installed.

In both maps, one node is sending 10 Interests per second and there exists one content source in the network that holds the content of size 1.752MB. We experimented with three different Interest Lifetime (I.L.) values, i.e. the time that the Interest is being forwarded through the network, before it expires. 
For the first and second path selection approach, we compare MMM-VNDN and iMMM-VNDN, to show the advantages that our new protocol offers. For the third path selection approach, we compare our protocol with the flooding strategy, which broadcasts every Internet message into the network, the best route strategy, which chooses the best face to forward an Interest according to its cost~\cite{afanasyev2016devGuide}, the NCC strategy~\cite{ncc}, the Content-Centric Vehicular Networking (CCVN) algorithm~\cite{amadeo2012upcomingVanets} and the Controlled Data and Interest packet propagation strategy (CODIE)~\cite{ahmed2016codie}. The presented results run for 149s and have a confidence interval of 95\%. Finally, for both of our proposed protocols we flood an Interest message with a broadcast MAC address every 10 seconds, as explained in Section~\ref{nodediscovery}. 

\subsection{Simulation Results}\label{simresults}
\subsubsection{Manhattan Map}
We compare the two routing protocols for the first path selection approach in \cref{scenario1isr}. iMMM-VNDN keeps the Intere-st Satisfaction Rate above 93\% compared to MMM-VNDN that keeps the ISR above 70\%. This is due to the fact that the decision of processing or discarding a message is performed in the strategy layer of the NDN stack, thus allowing us more control over the incoming messages. In addition, the average latency is lower for iMMM-VNDN than for MMM-VNDN and remains in between 2.5 - 3 ms, regardless of the number of nodes. iMMM-VNDN performs worse in terms of jitter, which stays almost at 0.5 ms independent of the density of the nodes. 

In the second presented path selection approach in~\cref{scenario2isr}, we observe that for both protocols the ISR fluctuates. Generally, iMMM-VNDN performs better than MMM- VNDN, but it is clear that choosing the path with the lowest latency does not guarantee that the path will be valid. In contrast, because of the path breaks, we see that the ISR for both approaches is lower than in~\cref{scenario1isr}. Also, despite the fact that the goal was to decrease the average latency, the average latency is kept at the same levels as in~\cref{scenario1isr}. The average jitter also remains the same for both algorithms, and we observe that there are small fluctuations in jitter for low ISR.
  
For the third path selection approach, we achieve the best results compared to the other two. This is due to the fact that the selection of a next hop is based on the average latency of the path together with the latest time when the path was established. The results for the ISR are shown in ~\cref{scenario3isr}. iMMM-VNDN achieves the highest ISR compared to other protocols, independent from the I.L. value. We also highlight that the ISR of MMM-VNDN strategy fluctuates from 70-80\% and is almost 10\% higher compared to flooding. In ~\cref{scenario3isr}, we observe that best-route  and CCVN strategies have higher ISR in some cases than MMM-VNDN. But, as shown in~\cref{scenario3ltn}, CCVN's delay of the delivered content is up to 20 times higher. 
    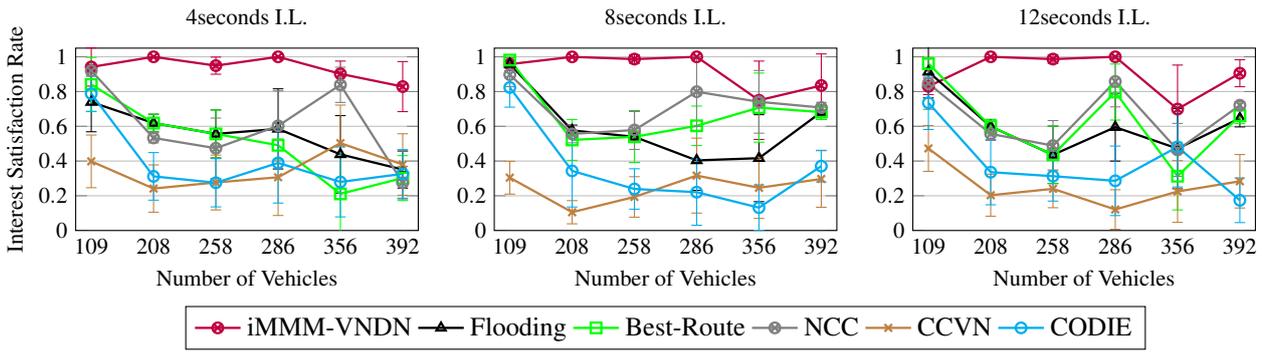
\begin{figure*}[!ht]
        	\centering
	\begin{tikzpicture}
 \begin{groupplot}[legend columns=7,legend to name= Fig10Legend,
            legend entries={{\normalsize iMMM-VNDN},{\normalsize Flooding},{\normalsize Best-Route},{\normalsize NCC},{\normalsize CCVN},{\normalsize CODIE}},
            group style={group size= 3 by 1, group name = fig10_plots},
	width  =0.35\textwidth,
	height = 4cm,
	major x tick style = transparent,
	ymajorgrids = true,
	xlabel = {Number of Vehicles},
	symbolic x coords={109,208,258,286,356,392},
	xtick = data,
	scaled y ticks = false,
	enlarge x limits=0.05,
	ymin=0,
	ymax=1.05,
     ytick={0,0.2,0.4,0.6,0.8,1},
    log ticks with fixed point={1000 sep=},
	ylabel near ticks	]
        \nextgroupplot[title=4seconds I.L.,ylabel={Interest Satisfaction Rate}]
     		]
  	\addplot[line width=0.30mm,
  	color=purple,mark=otimes,error bars/.cd, y dir=both, y explicit
  	]
  	plot coordinates {	
  (109, 0.941585) += (0,0.110003) -=(0,0.110003)
(208, 1.000000) += (0,0.000000) -=(0,0.000000)
(258, 0.949216) += (0,0.049236) -=(0,0.049236)
(286, 1.000000) += (0,0.000000) -=(0,0.000000)
(356, 0.901958) += (0,0.073533) -=(0,0.073533)
(392, 0.828584) += (0,0.144066) -=(0,0.144066)
  	};
  	\addplot[line width=0.30mm,
  	color=black,mark=triangle,
  	error bars/.cd, y dir=both, y explicit]
  	plot coordinates {	
  		(109, 0.739459) += (0,0.170878) -=(0,0.170878)
(208, 0.618697) += (0,0.050847) -=(0,0.050847)
(258, 0.556250) += (0,0.138241) -=(0,0.138241)
(286, 0.585126) += (0,0.230678) -=(0,0.230678)
(356, 0.438455) += (0,0.223021) -=(0,0.223021)
(392, 0.351003) += (0,0.106741) -=(0,0.106741)
  	};
  	\addplot[line width=0.30mm,
  	color=green,mark=square,
  	error bars/.cd, y dir=both, y explicit] 
  	plot coordinates {			
 (109, 0.840556) += (0,0.156117) -=(0,0.156117)
(208, 0.618697) += (0,0.050847) -=(0,0.050847)
(258, 0.556250) += (0,0.138241) -=(0,0.138241)
(286, 0.490875) += (0,0.079393) -=(0,0.079393)
(356, 0.211168) += (0,0.216260) -=(0,0.216260)
(392, 0.302018) += (0,0.129152) -=(0,0.129152)
  	};
  	\addplot[line width=0.30mm,
  	color=gray,mark=otimes,
  	error bars/.cd, y dir=both, y explicit] 
  	plot coordinates {			
  		(109, 0.919704) += (0,0.030183) -=(0,0.030183)
(208, 0.533091) += (0,0.028372) -=(0,0.028372)
(258, 0.473707) += (0,0.173818) -=(0,0.173818)
(286, 0.602503) += (0,0.201508) -=(0,0.201508)
(356, 0.838767) += (0,0.101427) -=(0,0.101427)
(392, 0.275688) += (0,0.093936) -=(0,0.093936)
  	};
      \addplot[line width=0.30mm,
	color=brown,mark=x,
	error bars/.cd, y dir=both, y explicit,
	] plot coordinates {			
		(109, 0.398141) += (0,0.151578) -=(0,0.151578)
(208, 0.241722) += (0,0.136052) -=(0,0.136052)
(258, 0.276528) += (0,0.157826) -=(0,0.157826)
(286, 0.306648) += (0,0.220040) -=(0,0.220040)
(356, 0.502849) += (0,0.219349) -=(0,0.219349)
(392, 0.380116) += (0,0.176976) -=(0,0.176976)
	};
\addplot[line width=0.30mm,
	color=cyan,mark=o,
	error bars/.cd, y dir=both, y explicit,
	] plot coordinates {			
		(109, 0.788019) += (0,0.101507) -=(0,0.101507)
(208, 0.311815) += (0,0.137010) -=(0,0.137010)
(258, 0.275051) += (0,0.139031) -=(0,0.139031)
(286, 0.389251) += (0,0.231047) -=(0,0.231047)
(356, 0.279661) += (0,0.201172) -=(0,0.201172)
(392, 0.326110) += (0,0.139784) -=(0,0.139784)
	};
     \nextgroupplot[title=8seconds I.L.]
    \addplot[line width=0.30mm,
    color=purple,mark=otimes,error bars/.cd, y dir=both, y explicit
    ]
    plot coordinates {	
(109, 0.957581) += (0,0.037692) -=(0,0.037692)
(208, 0.999737) += (0,0.000516) -=(0,0.000516)
(258, 0.986906) += (0,0.024777) -=(0,0.024777)
(286, 1.000000) += (0,0.000000) -=(0,0.000000)
(356, 0.749781) += (0,0.225643) -=(0,0.225643)
(392, 0.834683) += (0,0.181955) -=(0,0.181955)
    };
    \addplot[line width=0.30mm,
    color=black,mark=triangle,error bars/.cd, y dir=both, y explicit
    ]
    plot coordinates {	
    	(109, 0.958345) += (0,0.022783) -=(0,0.022783)
(208, 0.576382) += (0,0.030407) -=(0,0.030407)
(258, 0.539462) += (0,0.147560) -=(0,0.147560)
(286, 0.403710) += (0,0.172493) -=(0,0.172493)
(356, 0.416530) += (0,0.250279) -=(0,0.250279)
(392, 0.681761) += (0,0.042486) -=(0,0.042486)
    };
     \addplot[line width=0.30mm,
    color=green,mark=square,
    error bars/.cd, y dir=both, y explicit,
    ] plot coordinates {			
    	(109, 0.981376) += (0,0.015368) -=(0,0.015368)
(208, 0.521429) += (0,0.117410) -=(0,0.117410)
(258, 0.539462) += (0,0.147560) -=(0,0.147560)
(286, 0.602860) += (0,0.113590) -=(0,0.113590)
(356, 0.707520) += (0,0.199006) -=(0,0.199006)
(392, 0.681761) += (0,0.042486) -=(0,0.042486)
    };
    \addplot[line width=0.30mm,
    color=gray,mark=otimes,
    error bars/.cd, y dir=both, y explicit,
    ] plot coordinates {			
    	(109, 0.897202) += (0,0.069469) -=(0,0.069469)
(208, 0.556088) += (0,0.051518) -=(0,0.051518)
(258, 0.578503) += (0,0.111599) -=(0,0.111599)
(286, 0.799066) += (0,0.200777) -=(0,0.200777)
(356, 0.740924) += (0,0.180438) -=(0,0.180438)
(392, 0.707817) += (0,0.029910) -=(0,0.029910)
    };
      \addplot[line width=0.30mm,
	color=brown,mark=x,
	error bars/.cd, y dir=both, y explicit,
	] plot coordinates {			
		(109, 0.304209) += (0,0.094502) -=(0,0.094502)
(208, 0.105429) += (0,0.066793) -=(0,0.066793)
(258, 0.193363) += (0,0.116504) -=(0,0.116504)
(286, 0.316232) += (0,0.215686) -=(0,0.215686)
(356, 0.245804) += (0,0.176197) -=(0,0.176197)
(392, 0.296761) += (0,0.162810) -=(0,0.162810)

	};
\addplot[line width=0.30mm,
	color=cyan,mark=o,
	error bars/.cd, y dir=both, y explicit,
	] plot coordinates {			
		(109, 0.823431) += (0,0.113267) -=(0,0.113267)
(208, 0.343222) += (0,0.207684) -=(0,0.207684)
(258, 0.239263) += (0,0.116355) -=(0,0.116355)
(286, 0.220754) += (0,0.190353) -=(0,0.190353)
(356, 0.131669) += (0,0.130814) -=(0,0.130814)
(392, 0.370675) += (0,0.091179) -=(0,0.091179)
	};
    \nextgroupplot[title=12seconds I.L.]	
      	 \addplot[line width=0.30mm,
      	 color=purple,mark=otimes,error bars/.cd, y dir=both, y explicit
      	 ]
      	 plot coordinates {	
     (109, 0.829850) += (0,0.047362) -=(0,0.047362)
(208, 1.000000) += (0,0.000000) -=(0,0.000000)
(258, 0.986906) += (0,0.024777) -=(0,0.024777)
(286, 1.000000) += (0,0.000000) -=(0,0.000000)
(356, 0.699047) += (0,0.253517) -=(0,0.253517)
(392, 0.905420) += (0,0.077549) -=(0,0.077549)
      	 };
      	 \addplot[line width=0.30mm,
      	 color=black,mark=triangle,error bars/.cd, y dir=both, y explicit
      	 ]
      	 plot coordinates {	
      	(109, 0.914121) += (0,0.150408) -=(0,0.150408)
(208, 0.598595) += (0,0.040706) -=(0,0.040706)
(258, 0.435653) += (0,0.165094) -=(0,0.165094)
(286, 0.595055) += (0,0.195273) -=(0,0.195273)
(356, 0.471566) += (0,0.230833) -=(0,0.230833)
(392, 0.645996) += (0,0.049503) -=(0,0.049503)
      	 };
      	 \addplot[line width=0.30mm,
      	 color=green,mark=square,
      	 error bars/.cd, y dir=both, y explicit,
      	 ] plot coordinates {			
  (109, 0.961036) += (0,0.025306) -=(0,0.025306)
(208, 0.598595) += (0,0.040706) -=(0,0.040706)
(258, 0.435653) += (0,0.165094) -=(0,0.165094)
(286, 0.797193) += (0,0.161177) -=(0,0.161177)
(356, 0.312926) += (0,0.194550) -=(0,0.194550)
(392, 0.660578) += (0,0.045232) -=(0,0.045232)
      	 };
      	 \addplot[line width=0.30mm,
      	 color=gray,mark=otimes,
      	 error bars/.cd, y dir=both, y explicit,
      	 ] plot coordinates {			
     (109, 0.846824) += (0,0.149075) -=(0,0.149075)
(208, 0.556343) += (0,0.037537) -=(0,0.037537)
(258, 0.488790) += (0,0.144300) -=(0,0.144300)
(286, 0.858128) += (0,0.146173) -=(0,0.146173)
(356, 0.463965) += (0,0.152661) -=(0,0.152661)
(392, 0.721090) += (0,0.020633) -=(0,0.020633)
      	 };
       \addplot[line width=0.30mm,
	color=brown,mark=x,
	error bars/.cd, y dir=both, y explicit,
	] plot coordinates {			
		(109, 0.472611) += (0,0.131972) -=(0,0.131972)
(208, 0.203268) += (0,0.120747) -=(0,0.120747)
(258, 0.239989) += (0,0.108708) -=(0,0.108708)
(286, 0.121035) += (0,0.113863) -=(0,0.113863)
(356, 0.224645) += (0,0.177619) -=(0,0.177619)
(392, 0.283504) += (0,0.153880) -=(0,0.153880)
	};
\addplot[line width=0.30mm,
	color=cyan,mark=o,
	error bars/.cd, y dir=both, y explicit,
	] plot coordinates {			
		(109, 0.735140) += (0,0.153433) -=(0,0.153433)
(208, 0.335918) += (0,0.187762) -=(0,0.187762)
(258, 0.312808) += (0,0.143380) -=(0,0.143380)
(286, 0.286122) += (0,0.200056) -=(0,0.200056)
(356, 0.481513) += (0,0.230610) -=(0,0.230610)
(392, 0.174121) += (0,0.128400) -=(0,0.128400)
	};
     \end{groupplot}
       legend style={
		at={(1,1.05)},
		anchor=south east,
		column sep=1ex}
     \node (fig10_Legend ) at ($(fig10_plots c2r1.center)-(0,2.5cm)$){\ref{Fig10Legend}};
       \end{tikzpicture}       
       \caption{Interest Satisfaction Rate in Luxembourg map for the 3rd proposed path selection approach}
       \label{scenario3isrlux}
    \end{figure*}

    \begin{figure*}[!ht]
        	\centering
	\begin{tikzpicture}
 \begin{groupplot}[legend columns=7,legend to name= Fig11Legend,
            legend entries={{\normalsize iMMM-VNDN},{\normalsize Flooding},{\normalsize Best-Route},{\normalsize NCC},{\normalsize CCVN},{\normalsize CODIE}},
            group style={group size= 3 by 1, group name = fig11_plots},
	width  =0.35\textwidth,
	height = 4cm,
	major x tick style = transparent,
	ymajorgrids = true,
	xlabel = {Number of Vehicles},
	symbolic x coords={109,208,258,286,356,392},
	xtick = data,
	scaled y ticks = false,
    ytick={2,1000,2000},
	enlarge x limits=0.05,
	ymin=-500,
	ymax=2500,
	ylabel near ticks	]
        \nextgroupplot[title=4seconds I.L.,ylabel={Average Latency (ms) }]
     		      	\addplot[line width=0.30mm,
  	color=purple,mark=otimes,error bars/.cd, y dir=both, y explicit
  	]
  	plot coordinates {	
  		(109, 2.761501) += (0,0.188414) -=(0,0.188414)
(208, 3.147267) += (0,0.340862) -=(0,0.340862)
(258, 2.614158) += (0,0.130244) -=(0,0.130244)
(286, 2.470516) += (0,0.010356) -=(0,0.010356)
(356, 2.811510) += (0,0.377803) -=(0,0.377803)
(392, 4.254087) += (0,1.911948) -=(0,1.911948)
  	};
  	\addplot[line width=0.30mm,
  	color=black,mark=triangle,
  	error bars/.cd, y dir=both, y explicit]
  	plot coordinates {	
  			(109, 1093.815100) += (0,181.761480) -=(0,181.761480)
(208, 1258.816000) += (0,16.869440) -=(0,16.869440)
(258, 1758.914000) += (0,44.423429) -=(0,44.423429)
(286, 1598.044000) += (0,48.680403) -=(0,48.680403)
(356, 1838.840000) += (0,312.820737) -=(0,312.820737)
(392, 1696.850000) += (0,146.055152) -=(0,146.055152)
  	};
  	\addplot[line width=0.30mm,
  	color=green,mark=square,
  	error bars/.cd, y dir=both, y explicit] 
  	plot coordinates {			
  			(109, 985.670100) += (0,202.653678) -=(0,202.653678)
(208, 1258.816000) += (0,16.869440) -=(0,16.869440)
(258, 1758.914000) += (0,44.423429) -=(0,44.423429)
(286, 1892.550000) += (0,129.497622) -=(0,129.497622)
(356, 1616.615000) += (0,158.318742) -=(0,158.318742)
(392, 1725.337000) += (0,181.589895) -=(0,181.589895)
  	};
  	\addplot[line width=0.30mm,
  	color=gray,mark=otimes,
  	error bars/.cd, y dir=both, y explicit] 
  	plot coordinates {			
  			(109, 917.582000) += (0,138.906441) -=(0,138.906441)
(208, 1348.073000) += (0,20.771726) -=(0,20.771726)
(258, 1914.313000) += (0,35.876226) -=(0,35.876226)
(286, 1778.336000) += (0,76.237914) -=(0,76.237914)
(356, 1884.406000) += (0,125.436642) -=(0,125.436642)
(392, 1825.724000) += (0,112.376201) -=(0,112.376201)
  	};
      \addplot[line width=0.30mm,
	color=brown,mark=x,
	error bars/.cd, y dir=both, y explicit,
	] plot coordinates {			
			(109, 1574.465000) += (0,178.607352) -=(0,178.607352)
(208, 1661.044000) += (0,101.701710) -=(0,101.701710)
(258, 1596.486000) += (0,181.507211) -=(0,181.507211)
(286, 1848.496000) += (0,321.498783) -=(0,321.498783)
(356, 1953.556000) += (0,200.328805) -=(0,200.328805)
(392, 1787.378000) += (0,219.970262) -=(0,219.970262)
	};
\addplot[line width=0.30mm,
	color=cyan,mark=o,
	error bars/.cd, y dir=both, y explicit,
	] plot coordinates {			
		(109, 1554.133000) += (0,124.605245) -=(0,124.605245)
(208, 1529.846000) += (0,63.869225) -=(0,63.869225)
(258, 1558.603000) += (0,235.809576) -=(0,235.809576)
(286, 1789.607000) += (0,188.119235) -=(0,188.119235)
(356, 1546.222000) += (0,86.967409) -=(0,86.967409)
(392, 1676.705000) += (0,465.157937) -=(0,465.157937)
	};
     \nextgroupplot[title=8seconds I.L.]
    \addplot[line width=0.30mm,
    color=purple,mark=otimes,error bars/.cd, y dir=both, y explicit
    ]
    plot coordinates {	
(109, 2.759876) += (0,0.238571) -=(0,0.238571)
(208, 3.351911) += (0,0.464845) -=(0,0.464845)
(258, 2.512893) += (0,0.028358) -=(0,0.028358)
(286, 2.494407) += (0,0.020463) -=(0,0.020463)
(356, 9.218022) += (0,5.851119) -=(0,5.851119)
(392, 31.975996) += (0,56.005974) -=(0,56.005974)
    };
    \addplot[line width=0.30mm,
    color=black,mark=triangle,error bars/.cd, y dir=both, y explicit
    ]
    plot coordinates {	
    		(109, 1015.994800) += (0,115.827688) -=(0,115.827688)
(208, 1270.062000) += (0,23.994974) -=(0,23.994974)
(258, 1656.972000) += (0,49.976986) -=(0,49.976986)
(286, 1584.601000) += (0,97.476143) -=(0,97.476143)
(356, 1813.670000) += (0,168.335495) -=(0,168.335495)
(392, 1497.832000) += (0,22.955514) -=(0,22.955514)
    };
     \addplot[line width=0.30mm,
    color=green,mark=square,
    error bars/.cd, y dir=both, y explicit,
    ] plot coordinates {			
    		(109, 759.503000) += (0,37.746087) -=(0,37.746087)
(208, 1144.622000) += (0,250.401076) -=(0,250.401076)
(258, 1656.972000) += (0,49.976986) -=(0,49.976986)
(286, 1867.388000) += (0,227.410215) -=(0,227.410215)
(356, 1674.033000) += (0,123.665665) -=(0,123.665665)
(392, 1497.832000) += (0,22.955514) -=(0,22.955514)
    };
    \addplot[line width=0.30mm,
       color=gray,mark=otimes,
    error bars/.cd, y dir=both, y explicit,
    ] plot coordinates {			
    	(109, 972.386100) += (0,68.962122) -=(0,68.962122)
(208, 1323.276000) += (0,21.619006) -=(0,21.619006)
(258, 1941.160000) += (0,43.744300) -=(0,43.744300)
(286, 1585.723000) += (0,118.886276) -=(0,118.886276)
(356, 1849.326000) += (0,249.536207) -=(0,249.536207)
(392, 1707.515000) += (0,47.574922) -=(0,47.574922)
    };
      \addplot[line width=0.30mm,
	color=brown,mark=x,
	error bars/.cd, y dir=both, y explicit,
	] plot coordinates {			
		(109, 1653.082000) += (0,110.790819) -=(0,110.790819)
(208, 1704.014000) += (0,190.346905) -=(0,190.346905)
(258, 1470.509800) += (0,307.515579) -=(0,307.515579)
(286, 1564.914500) += (0,333.316198) -=(0,333.316198)
(356, 1661.405500) += (0,444.433282) -=(0,444.433282)
(392, 2012.239000) += (0,459.629638) -=(0,459.629638)
	};
\addplot[line width=0.30mm,
	color=cyan,mark=o,
	error bars/.cd, y dir=both, y explicit,
	] plot coordinates {			
			(109, 1432.569800) += (0,152.760003) -=(0,152.760003)
(208, 1721.218000) += (0,446.659143) -=(0,446.659143)
(258, 1487.467000) += (0,90.828482) -=(0,90.828482)
(286, 1626.987000) += (0,136.287606) -=(0,136.287606)
(356, 1628.290000) += (0,327.841819) -=(0,327.841819)
(392, 1215.116300) += (0,182.399724) -=(0,182.399724)
	};
    \nextgroupplot[title=12seconds I.L.]	
             	 \addplot[line width=0.30mm,
      	 color=purple,mark=otimes,error bars/.cd, y dir=both, y explicit
      	 ]
          plot coordinates {	
      	 	(109, 4.018254) += (0,0.886470) -=(0,0.886470)
(208, 3.229508) += (0,0.385338) -=(0,0.385338)
(258, 2.512893) += (0,0.028358) -=(0,0.028358)
(286, 2.494407) += (0,0.020463) -=(0,0.020463)
(356, 6.366373) += (0,4.663930) -=(0,4.663930)
(392, 5.544711) += (0,3.886378) -=(0,3.886378)
      	 };
      	 \addplot[line width=0.30mm,
      	 color=black,mark=triangle,error bars/.cd, y dir=both, y explicit
      	 ]
      	 plot coordinates {	
      		(109, 839.148700) += (0,113.346073) -=(0,113.346073)
(208, 1285.691000) += (0,30.966252) -=(0,30.966252)
(258, 1629.993000) += (0,49.663418) -=(0,49.663418)
(286, 1419.362000) += (0,56.374245) -=(0,56.374245)
(356, 1633.198000) += (0,169.086392) -=(0,169.086392)
(392, 1490.441000) += (0,40.451888) -=(0,40.451888)
      	 };
      	 \addplot[line width=0.30mm,
      	 color=green,mark=square,
      	 error bars/.cd, y dir=both, y explicit,
      	 ] plot coordinates {			
      		(109, 869.619400) += (0,157.735468) -=(0,157.735468)
(208, 1285.691000) += (0,30.966252) -=(0,30.966252)
(258, 1629.993000) += (0,49.663418) -=(0,49.663418)
(286, 1468.311000) += (0,109.041337) -=(0,109.041337)
(356, 1867.863000) += (0,168.221432) -=(0,168.221432)
(392, 1489.540000) += (0,40.277963) -=(0,40.277963)
      	 };
      	 \addplot[line width=0.30mm,
      	 color=gray,mark=otimes,
      	 error bars/.cd, y dir=both, y explicit,
      	 ] plot coordinates {			
      	 	(109, 1182.809900) += (0,189.840207) -=(0,189.840207)
(208, 1370.126000) += (0,30.831442) -=(0,30.831442)
(258, 1911.741000) += (0,37.408354) -=(0,37.408354)
(286, 1676.664000) += (0,74.725619) -=(0,74.725619)
(356, 2098.869000) += (0,137.700365) -=(0,137.700365)
(392, 1710.745000) += (0,42.540637) -=(0,42.540637)
      	 };
       \addplot[line width=0.30mm,
	color=brown,mark=x,
	error bars/.cd, y dir=both, y explicit,
	] plot coordinates {			
			(109, 1410.256900) += (0,211.203284) -=(0,211.203284)
(208, 1445.216400) += (0,334.129434) -=(0,334.129434)
(258, 1487.856000) += (0,190.803849) -=(0,190.803849)
(286, 1575.997300) += (0,402.101034) -=(0,402.101034)
(356, 1442.018900) += (0,360.115674) -=(0,360.115674)
(392, 1653.041900) += (0,312.071742) -=(0,312.071742)
	};
\addplot[line width=0.30mm,
	color=cyan,mark=o,
	error bars/.cd, y dir=both, y explicit,
	] plot coordinates {			
			(109, 1328.704900) += (0,181.686272) -=(0,181.686272)
(208, 1509.075000) += (0,35.957167) -=(0,35.957167)
(258, 1617.481000) += (0,270.913615) -=(0,270.913615)
(286, 1699.852000) += (0,160.334962) -=(0,160.334962)
(356, 1512.412000) += (0,132.315373) -=(0,132.315373)
(392, 1582.122000) += (0,161.480239) -=(0,161.480239)
	};
     \end{groupplot}
       legend style={
		at={(1,1.05)},
		anchor=south east,
		column sep=1ex}
     \node (fig11_Legend ) at ($(fig11_plots c2r1.center)-(0,2.5cm)$){\ref{Fig11Legend}};
       \end{tikzpicture}       
       \caption{Average Latency in Luxembourg map for the 3rd proposed path selection approach}
       \label{scenario3ltnlux}
    \end{figure*}
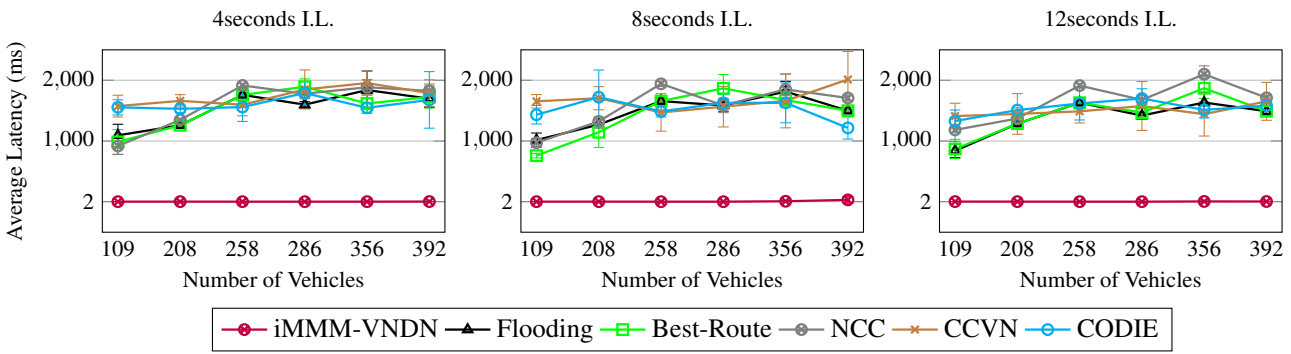

      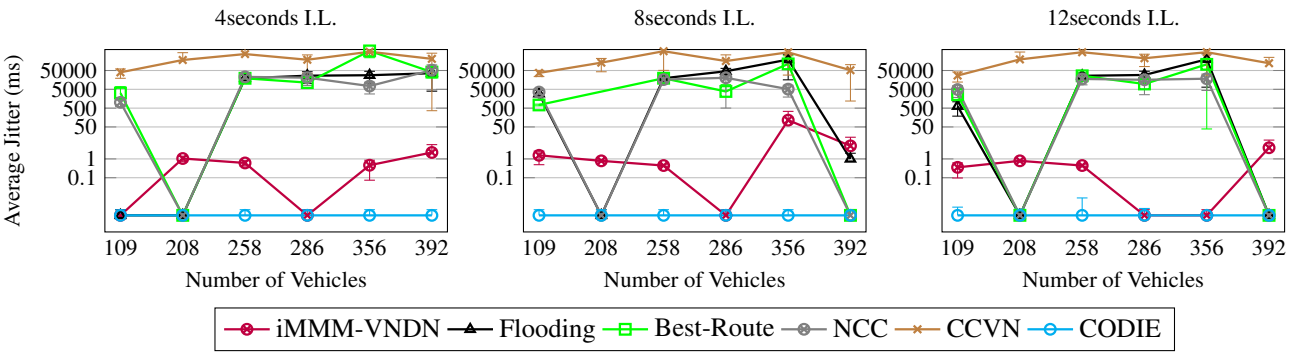
\begin{figure*}[!ht]
       	\centering
       	\begin{tikzpicture}
       	\begin{groupplot}[legend columns=7,legend to name= Fig12Legend,
       	legend entries={{\normalsize iMMM-VNDN},{\normalsize Flooding},{\normalsize Best-Route},{\normalsize NCC},{\normalsize CCVN},{\normalsize CODIE}},
       	group style={group size= 3 by 1, group name = fig12_plots},
       	width  =0.35\textwidth,
       	height = 4cm,
       	major x tick style = transparent,
       	ymajorgrids = true,
       	xlabel = {Number of Vehicles},
       	symbolic x coords={109,208,258,286,356,392},
       	xtick = data,
       	scaled y ticks = false,
       	enlarge x limits=0.05,
       	ymin=-0.1,
       	ymax=650500,
       	ymode=log,
       	ytick={0,0.1,1,50,500,5000,50000},
       	log ticks with fixed point={1000 sep=},
       	ylabel near ticks	]
       	\nextgroupplot[title=4seconds I.L.,ylabel={Average Jitter (ms) }]
       	]
       	\addplot[line width=0.30mm,
       	color=purple,mark=otimes,error bars/.cd, y dir=both, y explicit
       	]
       	plot coordinates {	
       		(109, 0.001) += (0,0.001) -=(0,0.001)
(208, 1.050622) += (0,0.344584) -=(0,0.344584)
(258, 0.609809) += (0,0.192160) -=(0,0.192160)
(286, 0.001) += (0,0.001) -=(0,0.001)
(356, 0.461734) += (0,0.387873) -=(0,0.387873)
(392, 2.198646) += (0,3.538861) -=(0,3.538861)
       	};
       	\addplot[line width=0.30mm,
       	color=black,mark=triangle,
       	error bars/.cd, y dir=both, y explicit]
       	plot coordinates {	
       		(109, 0.001) += (0,0.001) -=(0,0.001)
(208, 0.001) += (0,0.001) -=(0,0.001)
(258, 19284.650000) += (0,8429.594195) -=(0,8429.594195)
       		(286, 26516) += (0,16434.787070) -=(0,16434.787070)
       		(356, 28042.93) += (0,17381.188089) -=(0,17381.188089)
(392, 35856.460000) += (0,31875.849036) -=(0,31875.849036)
       	};
       	\addplot[line width=0.30mm,
       	color=green,mark=square,
       	error bars/.cd, y dir=both, y explicit] 
       	plot coordinates {			
      (109, 3163.639100) += (0,3756.484208) -=(0,3756.484208)
(208, 0.001) += (0,0.001) -=(0,0.001)
(258, 19284.650000) += (0,8429.594195) -=(0,8429.594195)
(286, 11369.220000) += (0,14863.292498) -=(0,14863.292498)
(356, 550883.346000) += (0,311267.958073) -=(0,311267.958073)
(392, 41259.322000) += (0,20968.544907) -=(0,20968.544907)
       	};
       	\addplot[line width=0.30mm,
       	color=gray,mark=otimes,
       	error bars/.cd, y dir=both, y explicit] 
       	plot coordinates {			
       		(109, 1013.397200) += (0,410.729378) -=(0,410.729378)
            (208, 0.001) += (0,0.001) -=(0,0.001)
             (258, 22913.080000) += (0,9463.104767) -=(0,9463.104767)
       		(286, 20798.21) += (0,12890.733952) -=(0,12890.733952)
       		(356, 7450.3) += (0,4617.743781) -=(0,4617.743781)
       		(392, 52256.020000) += (0,47544.948455) -=(0,47544.948455)
       	};
       	\addplot[line width=0.30mm,
       	color=brown,mark=x,
       	error bars/.cd, y dir=both, y explicit,
       	] plot coordinates {			
       		(109, 39941.248000) += (0,20665.293640) -=(0,20665.293640)
(208, 180919.594000) += (0,262885.972168) -=(0,262885.972168)
(258, 374248.420000) += (0,458177.085842) -=(0,458177.085842)
(286, 186629.297000) += (0,151018.008642) -=(0,151018.008642)
(356, 496537.410000) += (0,929857.938739) -=(0,929857.938739)
(392, 208114.170000) += (0,207748.801360) -=(0,207748.801360)
       	};
       	\addplot[line width=0.30mm,
       	color=cyan,mark=o,
       	error bars/.cd, y dir=both, y explicit,
       	] plot coordinates {			
       		(109, 0.001) += (0,0.001) -=(0,0.001)
       		(208, 0.001) += (0,0.001) -=(0,0.001)
       		(258, 0.001) += (0,0.001) -=(0,0.001)
       		(286, 0.001) += (0,0.001) -=(0,0.001)
       		(356, 0.001) += (0,0.001) -=(0,0.001)
       		(392, 0.001) += (0,0.001) -=(0,0.001)
       	};
           	\nextgroupplot[title=8seconds I.L.]
       	\addplot[line width=0.30mm,
       	color=purple,mark=otimes,error bars/.cd, y dir=both, y explicit
       	]
       	plot coordinates {	
   (109, 1.551155) += (0,1.058740) -=(0,1.058740)
(208, 0.789264) += (0,0.084939) -=(0,0.084939)
(258, 0.444834) += (0,0.042852) -=(0,0.042852)
(286, 0.001) += (0,0.001) -=(0,0.001)
(356, 114.767214) += (0,222.303695) -=(0,222.303695)
(392, 4.934208) += (0,9.274368) -=(0,9.274368)
       	};
       	\addplot[line width=0.30mm,
       	color=black,mark=triangle,error bars/.cd, y dir=both, y explicit
       	]
       	plot coordinates {	
       		(109, 2926.828000) += (0,1562.851237) -=(0,1562.851237)
(208, 0.001) += (0,0.001) -=(0,0.001)
(258, 19407.050000) += (0,8528.215078) -=(0,8528.215078)
(286, 45464.890000) += (0,19392.835253) -=(0,19392.835253)
(356, 190409.148000) += (0,174441.913713) -=(0,174441.913713)
(392, 001) += (0,001) -=(0,001)
       	};
       	\addplot[line width=0.30mm,
       	color=green,mark=square,
       	error bars/.cd, y dir=both, y explicit,
       	] plot coordinates {			
       		(109, 737.319400) += (0,73.813820) -=(0,73.813820)
(208, 0.000000) += (0,0.000000) -=(0,0.000000)
(258, 19407.050000) += (0,8528.215078) -=(0,8528.215078)
(286, 3837.390000) += (0,5016.354507) -=(0,5016.354507)
(356, 112285.898000) += (0,192400.676475) -=(0,192400.676475)
(392, 0.001) += (0,0.001) -=(0,0.001)
       	};
       	\addplot[line width=0.30mm,
       	color=gray,mark=otimes,
       	error bars/.cd, y dir=both, y explicit,
       	] plot coordinates {			
       		(109, 3574.310000) += (0,1517.410478) -=(0,1517.410478)
(208, 0.001) += (0,0.001) -=(0,0.001)
(258, 16475.830000) += (0,4975.065668) -=(0,4975.065668)
(286, 20456.689000) += (0,19953.207816) -=(0,19953.207816)
(356, 5107.866) += (0,3165.888146) -=(0,3165.888146)
(392, 0.001) += (0,0.001) -=(0,0.001)
       	};
       	\addplot[line width=0.30mm,
       	color=brown,mark=x,
       	error bars/.cd, y dir=both, y explicit,
       	] plot coordinates {			
       		(109, 37178.720000) += (0,13402.343155) -=(0,13402.343155)
(208, 130710.135000) += (0,87756.299566) -=(0,87756.299566)
(258, 529889.430000) += (0,479101.007110) -=(0,479101.007110)
(286, 159122.525000) += (0,180121.207509) -=(0,180121.207509)
(356, 462835.442000) += (0,434289.308229) -=(0,434289.308229)
(392, 51872.731000) += (0,50680.179984) -=(0,50680.179984)
       	};
       	\addplot[line width=0.30mm,
       	color=cyan,mark=o,
       	error bars/.cd, y dir=both, y explicit,
       	] plot coordinates {			
       		(109, 0.001) += (0,0.001) -=(0,0.001)
       		(208, 0.001) += (0,0.001) -=(0,0.001)
       		(258, 0.001) += (0,0.001) -=(0,0.001)
       		(286, 0.001) += (0,0.001) -=(0,0.001)
       		(356, 0.001) += (0,0.001) -=(0,0.001)
       		(392, 0.001) += (0,0.001) -=(0,0.001)
       	};
              	\nextgroupplot[title=12seconds I.L.]	
       	\addplot[line width=0.30mm,
       	color=purple,mark=otimes,error bars/.cd, y dir=both, y explicit
       	]
       	plot coordinates {	
       		(109, 0.356076) += (0,0.259204) -=(0,0.259204)
(208, 0.790802) += (0,0.073449) -=(0,0.073449)
(258, 0.444834) += (0,0.042852) -=(0,0.042852)
(286, 0.001) += (0,0.001) -=(0,0.001)
(356, 0.001) += (0,0.001) -=(0,0.001)
(392, 3.918070) += (0,6.093868) -=(0,6.093868)
       	};
       	\addplot[line width=0.30mm,
       	color=black,mark=triangle,error bars/.cd, y dir=both, y explicit
       	]
       	plot coordinates {	
       		(109, 637.710200) += (0,450.145174) -=(0,450.145174)
(208, 0.001) += (0,0.001) -=(0,0.001)
(258, 26201.960000) += (0,10603.798144) -=(0,10603.798144)
(286, 28939.421000) += (0,16699.546581) -=(0,16699.546581)
(356, 196793.177000) += (0,190285.326970) -=(0,190285.326970)
(392, 0.001) += (0,0.001) -=(0,0.001)

       	};
       	\addplot[line width=0.30mm,
       	color=green,mark=square,
       	error bars/.cd, y dir=both, y explicit,
       	] plot coordinates {			
       	(109, 2596.434500) += (0,2898.698071) -=(0,2898.698071)
(208, 0.001) += (0,0.001) -=(0,0.001)
(258, 26201.960000) += (0,10603.798144) -=(0,10603.798144)
(286, 9598.320000) += (0,10488.706343) -=(0,10488.706343)
(356, 106317.202000) += (0,106278.092971) -=(0,106278.092971)
(392, 0.001) += (0,0.001) -=(0,0.001)       	};
       	\addplot[line width=0.30mm,
       	       	color=gray,mark=otimes,
       	error bars/.cd, y dir=both, y explicit,
       	] plot coordinates {			
       		(109, 4867.66) += (0,3017.006925) -=(0,3017.006925)
       		(208, 0.001) += (0,0.001) -=(0,0.001)
(258, 17770.540000) += (0,9122.407868) -=(0,9122.407868)
(286, 16066.152000) += (0,13474.669793) -=(0,13474.669793)
(356, 18318.747000) += (0,13965.577080) -=(0,13965.577080)
(392, 0.001) += (0,0.001) -=(0,0.001)
       	};
       	\addplot[line width=0.30mm,
       	color=brown,mark=x,
       	error bars/.cd, y dir=both, y explicit,
       	] plot coordinates {			
       		(109, 26973.644000) += (0,14805.112278) -=(0,14805.112278)
(208, 194399.861000) += (0,285128.879487) -=(0,285128.879487)
(258, 467783.410000) += (0,523010.998735) -=(0,523010.998735)
(286, 224187.639000) += (0,142890.787146) -=(0,142890.787146)
(356, 471015.534000) += (0,442893.306884) -=(0,442893.306884)
(392, 122103.530000) += (0,126994.554138) -=(0,126994.554138)
       	};
       	\addplot[line width=0.30mm,
       	color=cyan,mark=o,
       	error bars/.cd, y dir=both, y explicit,
       	] plot coordinates {			
       		(109, 0.001) += (0,0.001733) -=(0,0.001733)
       		(208, 0.001) += (0,0.001129) -=(0,0.001129)
       		(258, 0.001) += (0,0.007433) -=(0,0.007433)
       		(286, 0.001) += (0,0.001326) -=(0,0.001326)
       		(356, 0.001) += (0,0.000000) -=(0,0.000000)
       		(392, 0.001) += (0,0.000000) -=(0,0.000000)
       	};
       	\end{groupplot}
       	legend style={
       		at={(1,1.05)},
       		anchor=south east,
       		column sep=1ex}
       	\node (fig12_Legend ) at ($(fig12_plots c2r1.center)-(0,2.5cm)$){\ref{Fig12Legend}};
       	\end{tikzpicture}       
       	\caption{Average Jitter in Luxembourg map for the 3rd proposed path selection approach}
       	\label{scenario3jtlux}
       \end{figure*}   
       
~\cref{scenario3ltn} shows the average latency of each strategy. The results indicate that our algorithms together with the best-route strategy have the lowest latency for all network sizes. The average latency fluctuates for MMM-VNDN from 3 to 5 ms and for iMMM-VNDN from 2.5 to 3 ms. This difference comes from the number of messages that are being delivered. Higher ISR for iMMM-VNDN means that more messages exist in the network. Thus, the possibility of collisions is higher. In contrast, other strategies achieve much higher delay than both of our proposed protocols. By considering the results of the ISR graph in \cref{scenario3isr} we manage to deliver more requested content. We reduced the latency by selecting paths. Furthermore, the network resources are released, because not all nodes participate in message transmissions, and the network is less congested. 

\cref{scenario3jt} shows the average jitter for all strategies. CODIE outperforms our protocols by keeping the jitter very low. iMMM-VNDN and MMM-VNDN perform similar to the previous path selection approaches, keeping the jitter above 1ms, compared to the other strategies.
\subsubsection{Luxembourg Map}
For the second set of experiments we used the map of the Luxembourg City Center, where we selected an area of 1km x 1km. There, we selected different time slots to extract the mobility traces.
Each simulation runs for 149 seconds.
We chose to use the traces that have more than 100 nodes to show what happens in a more dense network than the Manhattan map. Since with the Manhattan map iMMM-VNDN performs better than MMM-VNDN, we chose to include only iMMM-VNDN for these experiments in the 3rd path selection approach, i.e. when the selection of a path is based both on the newest creation time of the path combined with the lowest latency, compared to the other aforementioned strategies. \cref{scenario3isrlux} shows the ISR for all strategies. The ISR is kept almost the same for iMMM-VNDN independent of the I.L.. It is higher than 94\% for low number of nodes, and it decreases to around 80\%, if the node density is high. Moreover, iMMM-VNDN outperforms all other approaches considering the average latency, as seen in~\cref{scenario3ltnlux}. In particular, the average latency is stable at 2-4 ms, compared to all other strategies that have a latency bigger than 950 ms. We observe in~\cref{scenario3jtlux} that the average jitter of iMMM-VNDN is higher than for other approaches, in particular than for the CODIE strategy. This happens, because in a dense network CODIE only manages to retrieve messages that are 1-hop away from the requester node, thus the jitter is non-existent. But, since, in iMMM-VNDN multihop communication is supported, the jitter is increased compared to 1-hop communication algorithms. By creating unicast paths, we are able to reduce overall message transmissions in the network, and thus avoid collisions that happen in the network. This leads to a more stable environment for message transmissions, since collisions and congestion in general are avoided. 

%% file: conclusion.tex
In this paper, we present an enhanced routing protocol based on our previous work. In particular, we develop a routing strategy by using \textbf{M}ultihop, \textbf{M}ultipath and \textbf{M}ultichannel communication in \textbf{V}ANETs that uses \textbf{NDN} as a communication paradigm. We present a new routing protocol of the forwarding strategy, called \textit{\textbf{i}mproved} \textbf{MMM-VNDN}, iMMM-VNDN. In iMMM-VNDN we extract the MAC addresses from the strategy layer of NDN, contrary to MMM-VNDN where we added two new fields in the NDN messages. The MAC addresses are used for node identification and assist us in performing the forwarding decisions. The main goal is to save network resources in a VANET. Our results show that we satisfy more Interest messages than other approaches, by keeping the delay and jitter to a minimum, and thus improving the QoS of VANETs. 

%% file: main.bbl
\begin{thebibliography}{10}

\bibitem{giordano2002mobile}
S.~Giordano et~al.
\newblock Mobile ad hoc networks.
\newblock {\em Handbook of wireless networks and mobile computing}, pages
  325--346, 2002.

\bibitem{hartenstein2008tutorial}
H.~Hartenstein and L.~P. Laberteaux.
\newblock A tutorial survey on vehicular ad hoc networks.
\newblock {\em IEEE Communications magazine}, 46(6), 2008.

\bibitem{chen2014vendnet}
M.~Chen, D.~O. Mau, Y.~Zhang, T.~Taleb, and V.~C. Leung.
\newblock Vendnet: Vehicular named data network.
\newblock {\em Vehicular Communications}, 1(4):208--213, 2014.

\bibitem{zhang2014named}
L.~Zhang, A.~Afanasyev, J.~Burke, V.~Jacobson, P.~Crowley, C.~Papadopoulos,
  L.~Wang, B.~Zhang, et~al.
\newblock Named data networking.
\newblock {\em ACM SIGCOMM Computer Communication Review}, 44(3):66--73, 2014.

\bibitem{kalogeiton2017multihop}
E.~Kalogeiton, T.~Kolonko, and T.~Braun.
\newblock A multihop and multipath routing protocol using ndn for vanets.
\newblock In {\em Ad Hoc Networking Workshop (Med-Hoc-Net), 2017 16th Annual
  Mediterranean}, pages 1--8. IEEE, 2017.

\bibitem{jiang2008ieee}
Daniel Jiang and Luca Delgrossi.
\newblock Ieee 802.11 p: Towards an international standard for wireless access
  in vehicular environments.
\newblock In {\em Vehicular Technology Conference, 2008. VTC Spring 2008.
  IEEE}, pages 2036--2040. IEEE, 2008.

\bibitem{codeca2015luxembourg}
Lara Codeca, Rapha{\"e}l Frank, and Thomas Engel.
\newblock Luxembourg sumo traffic (lust) scenario: 24 hours of mobility for
  vehicular networking research.
\newblock In {\em Vehicular Networking Conference (VNC), 2015 IEEE}, pages
  1--8. IEEE, 2015.

\bibitem{abedi2008enhancing}
O.~Abedi, M.~Fathy, and J.~Taghiloo.
\newblock Enhancing aodv routing protocol using mobility parameters in vanet.
\newblock In {\em Computer Systems and Applications, 2008. AICCSA 2008.
  IEEE/ACS International Conference on}, pages 229--235. IEEE, 2008.

\bibitem{perkins2003ad}
C.~Perkins, E.~Belding-Royer, and S.~Das.
\newblock Ad hoc on-demand distance vector (aodv) routing.
\newblock RFC 3561, July 2003.

\bibitem{toutouh2012intelligent}
J.~Toutouh, J.~Garc{\'\i}a-Nieto, and E.~Alba.
\newblock Intelligent olsr routing protocol optimization for vanets.
\newblock {\em IEEE transactions on vehicular technology}, 61(4):1884--1894,
  2012.

\bibitem{clausen2003optimized}
T.~Clausen and P.~Jacquet.
\newblock Optimized link state routing protocol (olsr).
\newblock RFC 3626, October 2003.

\bibitem{luo2010new}
Y.~Luo, W.~Zhang, and Y.~Hu.
\newblock A new cluster based routing protocol for vanet.
\newblock In {\em Networks Security Wireless Communications and Trusted
  Computing (NSWCTC), 2010 Second International Conference on}, volume~1, pages
  176--180. IEEE, 2010.

\bibitem{saians2017efficient}
J.V. Sai{\'a}ns-V{\'a}zquez, M.~L{\'o}pez-Nores, Y.~Blanco-Fern{\'a}ndez, E.~F.
  Ord{\'o}{\~n}ez-Morales, J.~F. Bravo-Torres, and J.~J. Pazos-Arias.
\newblock Efficient and viable intersection-based routing in vanets on top of a
  virtualization layer.
\newblock {\em Annals of Telecommunications}, pages 1--12, 2017.

\bibitem{abdou2015priority}
W.~Abdou, B.~Darties, and N.~Mbarek.
\newblock Priority levels based multi-hop broadcasting method for vehicular ad
  hoc networks.
\newblock {\em Aannals of Telecommunications}, 70(7-8):359--368, 2015.

\bibitem{arjunwadkar2014introduction}
D.~P. Arjunwadkar.
\newblock Introduction of ndn with comparison to current internet architecture
  based on tcp/ip.
\newblock {\em International Journal of Computer Applications}, 105(5):31--35,
  November 2014.

\bibitem{amadeo2015forwarding}
M.~Amadeo, C.~Campolo, and A.~Molinaro.
\newblock Forwarding strategies in named data wireless ad hoc networks.
\newblock {\em J. Netw. Comput. Appl.}, 50(C):148--158, April 2015.

\bibitem{grassi2014vanet}
G.~Grassi, D.~Pesavento, G.~Pau, R.~Vuyyuru, R.~Wakikawa, and L.~Zhang.
\newblock Vanet via named data networking.
\newblock In {\em 2014 IEEE Conference on Computer Communications Workshops
  (INFOCOM WKSHPS)}, pages 410--415, April 2014.

\bibitem{anastasiades2016dynamic}
C.~Anastasiades, J.~Weber, and T.~Braun.
\newblock Dynamic unicast: Information-centric multi-hop routing for mobile
  ad-hoc networks.
\newblock {\em Computer Networks}, 107:208--219, 2016.

\bibitem{Amadeo2013enhancing}
M.~Amadeo, C.~Campolo, and A.~Molinaro.
\newblock Enhancing content-centric networking for vehicular environments.
\newblock {\em Computer Networks}, 57(16):3222 -- 3234, 2013.

\bibitem{amadeo2012contentcentric}
M.~Amadeo, C.~Campolo, and A.~Molinaro.
\newblock Content-centric vehicular networking: An evaluation study.
\newblock In {\em 2012 Third International Conference on The Network of the
  Future (NOF)}, pages 1--5, Nov 2012.

\bibitem{amadeo2012upcomingVanets}
M.~Amadeo, C.~Campolo, and A.~Molinaro.
\newblock Content-centric networking: is that a solution for upcoming vehicular
  networks?
\newblock In {\em Proceedings of the ninth ACM international workshop on
  Vehicular inter-networking, systems, and applications}, pages 99--102. ACM,
  2012.

\bibitem{jacobson2009networking}
V.~Jacobson, D.~K. Smetters, J.~D. Thornton, M.~F. Plass, N.~H. Briggs, and
  R.~L. Braynard.
\newblock Networking named content.
\newblock In {\em Proceedings of the 5th international conference on Emerging
  networking experiments and technologies}, pages 1--12. ACM, 2009.

\bibitem{893287}
Ieee guide for wireless access in vehicular environments (wave) - architecture.
\newblock {\em IEEE Std. 1609.0}, 2013.

\bibitem{ahmed2016codie}
Syed~Hassan Ahmed, Safdar~Hussain Bouk, Muhammad~Azfar Yaqub, Dongkyun Kim,
  Houbing Song, and Jaime Lloret.
\newblock Codie: Controlled data and interest evaluation in vehicular named
  data networks.
\newblock {\em IEEE Transactions on Vehicular Technology}, 65(6):3954--3963,
  2016.

\bibitem{gomes2017addressing}
J.~M. Gomes~Duarte, T.~Braun, and L.~Villas.
\newblock Addressing the effects of low vehicle densities in highly mobile
  vehicular named-data networks.
\newblock 2017.

\bibitem{ncc}
{NCC Strategy}.
\newblock \url{https://redmine.named-data.net/projects/nfd/wiki/CcndStrategy}.
\newblock Accessed September 2017.

\bibitem{kalogeiton2017sdn}
E.~Kalogeiton, Z.~Zhao, and T.~Braun.
\newblock Is sdn the solution for ndn-vanets?
\newblock In {\em Ad Hoc Networking Workshop (Med-Hoc-Net), 2017 16th Annual
  Mediterranean}, pages 1--6. IEEE, 2017.

\bibitem{duarte2017multi}
J.~M. Gomes~Duarte, E.~Kalogeiton, R.~Soua, G.~Manzo, M.R. Palattella,
  A.~Di~Maio, T.~Braun, T.~Engel, L.~Villas, and G.~Rizzo.
\newblock A multi-pronged approach to adaptive and context aware content
  dissemination in vanets.
\newblock {\em Mobile Networks and Applications}, pages 1--13, 2017.

\bibitem{mir2014lte}
Zeeshan~Hameed Mir and Fethi Filali.
\newblock Lte and ieee 802.11 p for vehicular networking: a performance
  evaluation.
\newblock {\em EURASIP Journal on Wireless Communications and Networking},
  2014(1):89, 2014.

\bibitem{karp2000gpsr}
Brad Karp and Hsiang-Tsung Kung.
\newblock Gpsr: Greedy perimeter stateless routing for wireless networks.
\newblock In {\em Proceedings of the 6th annual international conference on
  Mobile computing and networking}, pages 243--254. ACM, 2000.

\bibitem{rfc1889}
R.~Frederick, S.~L. Casner, V.~Jacobson, and H.~Schulzrinne.
\newblock {RTP: A Transport Protocol for Real-Time Applications}.
\newblock RFC 1889, January 1996.

\bibitem{afanasyev2012ndnSIM}
A.~Afanasyev, I.~Moiseenko, and L.~Zhang.
\newblock ndnsim: Ndn simulator for ns-3, October 2012.

\bibitem{mastorakis2016ndnSIM}
S.~Mastorakis, A.~Afanasyev, I.~Moiseenko, and L.~Zhang.
\newblock ndnsim 2: An updated ndn simulator for ns-3, November 2016.

\bibitem{afanasyev2016devGuide}
A.~Afanasyev, J.~Shi, B.~Zhang, L.~Zhang, et~al.
\newblock Nfd developer's guide, October 2016.

\bibitem{networkSimulator}
{The ns-3 Network Simulator}.
\newblock \url{https://www.nsnam.org}.
\newblock Accessed September 2017.

\bibitem{behrisch2011sumo}
M.~Behrisch, L.~Bieker, J.~Erdmann, and D.~Krajzewicz.
\newblock Sumo--simulation of urban mobility: an overview.
\newblock In {\em Proceedings of SIMUL 2011, The Third International Conference
  on Advances in System Simulation}. ThinkMind, 2011.

\end{thebibliography}
